\documentclass[aps,preprint,showpacs,nofootinbib,longbibliography]{revtex4-1}

\usepackage{amsmath} 
\usepackage{graphicx} 
\usepackage{amsthm}
\usepackage{amssymb} 
\usepackage{yfonts}
\usepackage{hyperref}

\newcommand{\dtilde}{\tilde{d}}
\newcommand{\mtilde}{\tilde{m}}
\newcommand{\mbar}{\bar{m}}
\newcommand{\wbar}{\bar{\omega}}
\newcommand{\qbar}{\bar{q}}
\newcommand{\Zbar}{\bar{Z}}
\newcommand{\Ebar}{\bar{E}}
\newcommand{\mutilde}{\tilde{\mu}}
\newcommand{\mgot}{\textswab{m}}

\begin{document}
\title{Holographic zero sound at finite temperature}
\author{Richard A.~Davison}
\email{r.davison1@physics.ox.ac.uk}
\author{Andrei O.~Starinets}
\email{andrei.starinets@physics.ox.ac.uk}
\affiliation{Rudolf Peierls
Centre for Theoretical Physics,
University of Oxford, 1 Keble Road, Oxford OX1 3NP, United Kingdom\\
\vspace{1cm}
}
\preprint{OUTP-11-53P}

\begin{abstract}
We use gauge-gravity duality to study the temperature dependence of the zero sound mode and the fundamental matter diffusion 
mode in the strongly  coupled 
 ${\cal N}=4$ $SU(N_c)$ supersymmetric Yang-Mills theory with $N_f$  ${\cal N}=2$ 
hypermultiplets in the $N_c\gg 1$, $N_c \gg N_f$ limit, which is holographically realized via the D3/D7 
brane system.  
In the high density limit $\mu \gg T$, three regimes can be identified in the behavior of these modes, analogous to the collisionless quantum, collisionless thermal and hydrodynamic regimes of a Landau Fermi-liquid. The transitions between 
the three regimes are characterized by the parameters $T/\mu$ and $(T/\mu)^2$ respectively, and in each
of these regimes the modes have a distinctively different temperature and momentum dependence. 
The collisionless-hydrodynamic transition occurs when the zero sound poles of the density-density correlator in the complex 
frequency plane collide on the imaginary axis to produce a hydrodynamic diffusion pole. 
We observe that  the properties characteristic of a Landau Fermi-liquid zero sound mode are present in the D3/D7 system 
despite the atypical  $T^6/\mu^3$ temperature scaling of the specific heat and an apparent lack of a directly identifiable Fermi surface. 
\end{abstract}

\maketitle

\thispagestyle{empty}

\section{Introduction}

The AdS/CFT correspondence
\cite{Maldacena:1997re,Gubser:1998bc,Witten:1998qj,Aharony:1999ti}  and, more generally, gauge-gravity
duality have been used extensively for studies of the thermodynamics 
and transport properties of strongly-interacting quantum field
theories at finite temperature and density. Applications of the
correspondence to real-world physics initially concentrated on QCD and,
in particular, the quark-gluon plasma \cite{Son:2007vk,CasalderreySolana:2011us}, but in recent
years there has been a surge of interest in using these tools to study
the physics of condensed matter systems  (see
e.g. \cite{Hartnoll:2009sz,Herzog:2009xv,McGreevy:2009xe} for an introduction to the
field).

In this paper we investigate the behaviour of the collective modes of the ${\cal N}=4$ 
$SU(N_c)$ supersymmetric Yang-Mills theory coupled to $N_f$  ${\cal N}=2$ fundamental hypermultiplets 
at infinitely large 't Hooft coupling $\lambda=g^2_{YM}N_c$ 
in the limit $N_c\gg 1$, $N_c \gg N_f$ and at finite temperature $T$ and finite density of the fundamental matter $d$.  
The holographic dual of this theory is provided by embedding $N_f$ D7-branes in the gravitational background created by $N_c$ D3-branes and treating the D7-branes as 
probes \cite{Karch:2002sh}.
Although a seemingly elaborate construction, this is one of
the simplest known finite-density field theories for which an explicit
dual is known, and it has some interesting properties.

At strictly zero temperature and zero hypermultiplet mass,
the theory supports a collective excitation which appears as a pole in the 
density-density correlator \cite{Karch:2008fa}. 
Such an excitation, found in \cite{Karch:2008fa} using the dual gravity methods and thus referred to here as 
the ``holographic zero sound", is reminiscent of the zero sound mode predicted to exist by 
Landau in a class of  Fermi liquid systems \cite{landau1} (see
also \cite{landaulifshitz,landaulifshitz10,qfttext,pinesnozieres,zerosoundreview}) and subsequently observed in liquid Helium-3 \cite{Abel:1966zz,dobbs}. In Landau Fermi-liquids (LFLs), 
the zero sound mode arises due to (non-thermal) interactions between the constituent
fermions which result in oscillations of the Fermi surface. The holographic 
zero sound mode in the D3/D7 system at zero temperature has a speed equal to that of the
ordinary (first) sound and an attentuation
proportional to the square of the momentum. This is identical to what one finds in 
the Fermi liquid models where the interaction strength 
(parameterised by the Fermi liquid coefficient $F_0$) 
approaches infinity \cite{pinesnozieres}. At the same time, the heat capacity of the D3/D7 system is proportional to $T^6$ at low temperatures 
whereas in normal Fermi liquids it is proportional to $T$.

Similar investigations have been made in other
string-theoretic constructions - in \cite{Kulaxizi:2008kv} it was
shown that the holographic zero sound mode persists at $T=0$ when the hypermultiplet is given a finite mass,
in \cite{Kulaxizi:2008jx} a similar mode was found in the $T=0$
D4/D8/$\overline{\text{D8}}$ theory at finite density, in
\cite{Hung:2009qk} a ``zero sound" mode was reported to exist in the
(1+1)-dimensional theory on the D3/D3 intersection and in \cite{Bergman:2011rf} a low temperature zero sound mode was found in the theory on the (2+1)-dimensional D3/D7 brane intersection. In the D4/D8/$\overline{\text{D8}}$ case - where the only
fundamental matter present is fermions - the heat capacity is proportional to 
$T$ as expected for a Landau Fermi-liquid. However, the imaginary part
of the zero sound mode in this case has an unconventional $q^3$
dependence, which is at odds with the predictions of LFL
theory. Holographic zero sound modes have also been discovered in
theories consisting of probe DBI actions embedded in Lifshitz
(i.e. non-relativistic) spacetimes
\cite{HoyosBadajoz:2010kd,Lee:2010ez}. Such modes
have a density-dependent speed, and their properties depend upon the
critical exponent of the spacetime. In addition to these, the field theory dual to (3+1)-dimensional Einstein-Maxwell gravity with a cosmological constant was found to support a long-lived sound mode at zero temperature \cite{Edalati:2010pn}.
\begin{figure*}
\centering
\includegraphics[scale=1.0]{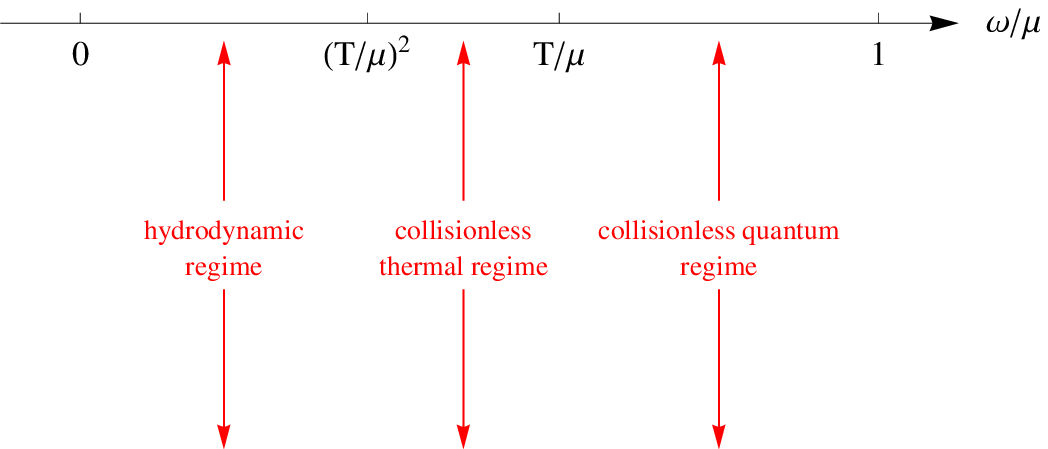}
\vspace{-10pt}
\includegraphics[scale=1.0]{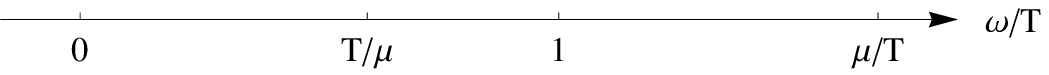}
\caption{{\small Relative scales in the hydrodynamic, collisionless thermal and collisionless 
quantum regimes of a Landau Fermi-liquid.}}
\label{arrow-scales}
\end{figure*}

A natural question to ask is what happens to the holographic zero sound mode when the temperature is turned on. In \cite{Kim:2008bv}, it was argued that at any infinitesimal temperature the zero sound mode of the D3/D7 theory is no longer dominant and instead density transport occurs mainly via diffusion. In this paper we show that this is not the case - at low temperatures, the holographic zero sound mode still dominates the density-density correlator with a dispersion relation very similar to that at $T=0$. Furthermore, we show (numerically) that the real part of this mode is independent of temperature whereas its imaginary part receives corrections proportional to $T^2$. At a sufficiently high temperature, there is a crossover to a hydrodynamic regime where the diffusion mode of \cite{Kim:2008bv} dominates (such crossovers are common - see \cite{Herzog:2007ij,Amado:2007yr,Amado:2008ji,Kaminski:2009dh,Kaminski:2009ce,Herzog:2010vz} for other holographic examples of this). This behaviour is familiar as it is the same as that predicted by Landau's theory of Fermi liquids, to which we now turn to help clarify these results and to identify the relevant scales involved.
\begin{table}[ht]
\caption{Relative scales in a Landau Fermi-liquid}
\begin{tabular}{|c|c|c|}
\hline
{} & $\hspace{1cm} \omega/\mu\,, q/\mu \;\;\; \mbox{variables}  \hspace{1cm}$  &    $\hspace{1cm}  \omega/T\,, q/T \;\;\; \mbox{variables} \hspace{1cm}$   \\ \hline 
\mbox{Hydrodynamic regime} &  $\frac{\omega}{\mu} \ll \left( \frac{T}{\mu} \right)^2$ &   $\frac{\omega}{T} \ll  \frac{T}{\mu} $ \\
\mbox{Collisionless thermal regime} &   $\left( \frac{T}{\mu} \right)^2 \ll
\frac{\omega}{\mu} \ll  \frac{T}{\mu}$      &   $ \frac{T}{\mu}  \ll \frac{\omega}{T} \ll  1$   \\
\mbox{Collisionless quantum regime}
 &    $\frac{T}{\mu}  \ll \frac{\omega}{\mu} \ll 1$     &      $1  \ll \frac{\omega}{T} \ll \frac{\mu}{T}$    \\ \hline
\end{tabular}
\label{tab0}
\end{table}

In Landau's theory of Fermi liquids, one assumes that the ground state is a degenerate system of interacting fermionic quasiparticles. The theory describes small fluctuations around this ground state due to 
quasiparticle-quasihole excitations and/or collective excitations, and remains valid as long as these fluctuations are sufficiently small. The applicability conditions of Landau Fermi-liquid theory are given by the inequalities
\begin{equation} \label{applicability}
T \ll \mu\,, \qquad \omega \ll \mu\,,
\end{equation}
where $\mu \sim d^{1/3}$ is the chemical potential. The first inequality guarantees that we are considering a degenerate (i.e. quantum) liquid, while the second ensures that the excitations are sufficiently macroscopic,
 i.e. their wavelengths are much larger than the characteristic interparticle distance 
(this condition also implies that the quasiparticles remain sufficiently close to the Fermi-surface:
$\omega \sim |\epsilon_p-\mu|\ll \mu$). The results of the 
Landau theory can be viewed as the leading order term in an expansion in 
powers of $\omega/\mu$ \cite{landaulifshitz,pinesnozieres}. 
\begin{figure*}
\centering
\includegraphics[scale=1.0]{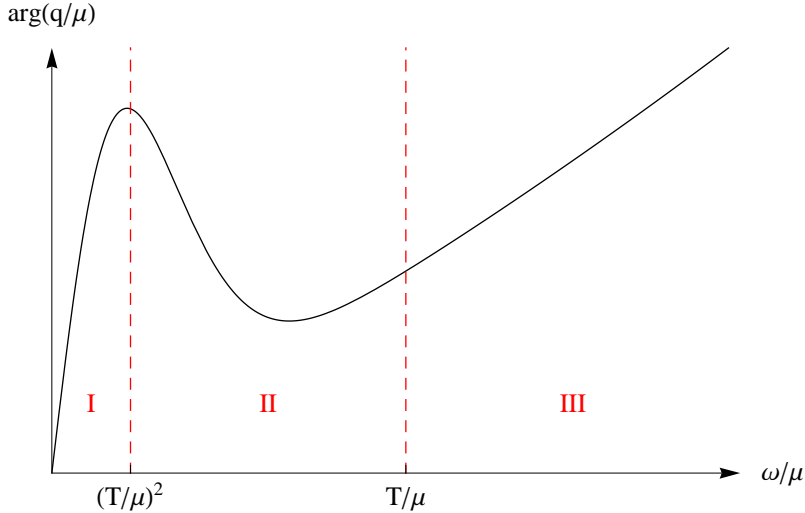}
\caption{{\small A sketch of the dependence of the sound mode damping on frequency in the hydrodynamic (I), 
collisionless thermal (II) and collisionless quantum (III) regimes of a Landau Fermi-liquid. 
First sound propagates in region I while the zero sound mode exists in regions II and III.}}
\label{argqplot}
\end{figure*}

In Fermi liquids, the zero sound mode is a longitudinal, gapless, collective excitation 
corresponding to oscillations of the Fermi surface around its equilibrium shape at zero temperature. 
Its dispersion relation 
$\omega(q) = v_s \, q - i \Gamma_\omega$ contains a non-zero damping due to the mode's 
decay into quasiparticle-quasihole pairs. It is often more convenient to 
work with real frequency and complex wave vector: $q(\omega) = \omega /v_s + i \Gamma_q$. 
The damping rate is conveniently characterised 
by considering the argument of $q$, $\mbox{arg} \, q (\omega) = \mbox{Im} \, q/\mbox{Re} \, q$, 
as a function of the frequency.

The zero sound mode persists at small, non-zero temperatures as well, but its properties are altered
 by the thermal collisions of the quasiparticles. Such collisions have a characteristic frequency $\nu \sim 1/\tau$, 
where $\tau$ is the mean time between quasiparticle collisions. One distinguishes three regimes:
the hydrodynamic regime, characterized by $\omega \ll \nu$, the collisionless thermal (classical) regime,
with $\omega \gg \nu$, $\omega \ll T$, and the collisionless quantum regime, $\omega \gg \nu$, $\omega \gg T$.
The zero temperature zero sound mode persists essentially unaltered in the collisionless quantum regime, where thermal excitations are too weak and infrequent to influence it. As the temperature is increased, however, thermal 
excitations change the attenuation of the zero sound mode giving it nontrivial temperature  dependence.
In the hydrodynamic regime, thermal excitations destroy the zero sound mode completely. However, these excitations support the ordinary hydrodynamic sound mode with viscous damping, and the thermal diffusion mode.

In the collisionless regime, the frequency $\nu$ 
can be computed from kinetic theory applied to Fermi liquids \cite{pinesnozieres,landaulifshitz,landaulifshitz10}:
\begin{equation} \label{freq-quasi}
\nu \sim \frac{\pi^2 T^2 + \omega^2}{\varepsilon_F (1+e^{-\omega/T})}\,,
\end{equation}
where $\varepsilon_F \sim \mu$. The decay rate of the zero sound mode in the collisionless regime is 
$\Gamma_q \sim \nu$ \cite{pinesnozieres}. In the quantum collisionless regime one has then 
$\Gamma_q \sim \omega^2/\mu$ and  $\mbox{arg} \, q (\omega) \sim \omega/\mu$ whereas 
in the thermal collisionless regime $\Gamma_q \sim T^2/\mu$ and  $\mbox{arg} \, q (\omega) \sim 
(T/\mu)^2 \mu/\omega$. Finally, in the hydrodynamic regime, the conditions 
$\omega \ll \nu$ and $\omega \ll T$ lead to
$\omega/\mu \ll (T/\mu)^2$ (the ``quantum" limit $\omega \gg T$, $\omega \ll \nu$ lies outside of the LFL 
applicability range (\ref{applicability})). The attenuation of hydrodynamic (first) sound is determined by the viscosity and is proportional to $\omega^2/T^2$  \cite{pinesnozieres,landaulifshitz,landaulifshitz10}.

The dimensionless variables which are most convenient for identifying the three regimes are $\omega/\mu$ and $q/\mu$.
In these variables, the regions I, II and III corresponding to the hydrodynamic, 
collisionless thermal and collisionless quantum regimes respectively, are 
separated by the scales $(T/\mu)^2$ and $T/\mu$. Alternatively, in the language of the ``traditional" 
hydrodynamic variables $\omega/T$ and $q/T$, the relevant scales are $T/\mu$, $1$ and $\mu/T$
 (see Table \ref{tab0} and Fig.~\ref{arrow-scales}).
\begin{table}[ht]
\caption{Sound attenuation coefficients in a Landau Fermi-liquid}
\begin{tabular}{|c|c|c|c|}
\hline
{} & $\hspace{1cm} \Gamma_\omega  \hspace{1cm}$  &    $\hspace{1cm} \Gamma_q   \hspace{1cm}$ &    $\hspace{1cm}  \mbox{Arg}\, q   \hspace{1cm}$  \\ \hline 
\mbox{Hydrodynamic regime} & $\left(\frac{\mu}{T}\right)^2 \frac{q^2}{\mu}$ & $\frac{\mu\, \omega^2}{T^2}$ &   $\left(\frac{\mu}{T}\right)^2 \frac{\omega}{\mu}$  \\
\mbox{Collisionless thermal regime} & $\frac{T^2}{\mu}$ &   $\frac{T^2}{\mu}$ & 
$\left(\frac{T}{\mu}\right)^2 \frac{\mu}{\omega}$ \\
\mbox{Collisionless quantum regime} & $\frac{q^2}{\mu}$ & $\frac{\omega^2}{\mu}$ & $\frac{\omega}{\mu}$ \\ \hline
\end{tabular}
\label{tab1}
\end{table}

\begin{figure*}
\centering
\includegraphics[scale=1.0]{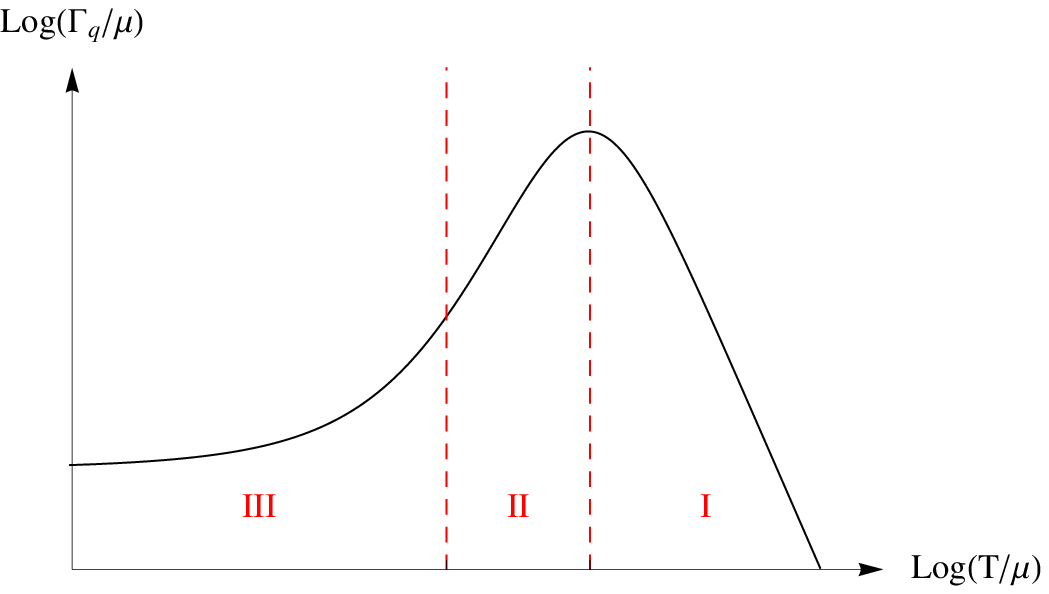}
\includegraphics[scale=0.5]{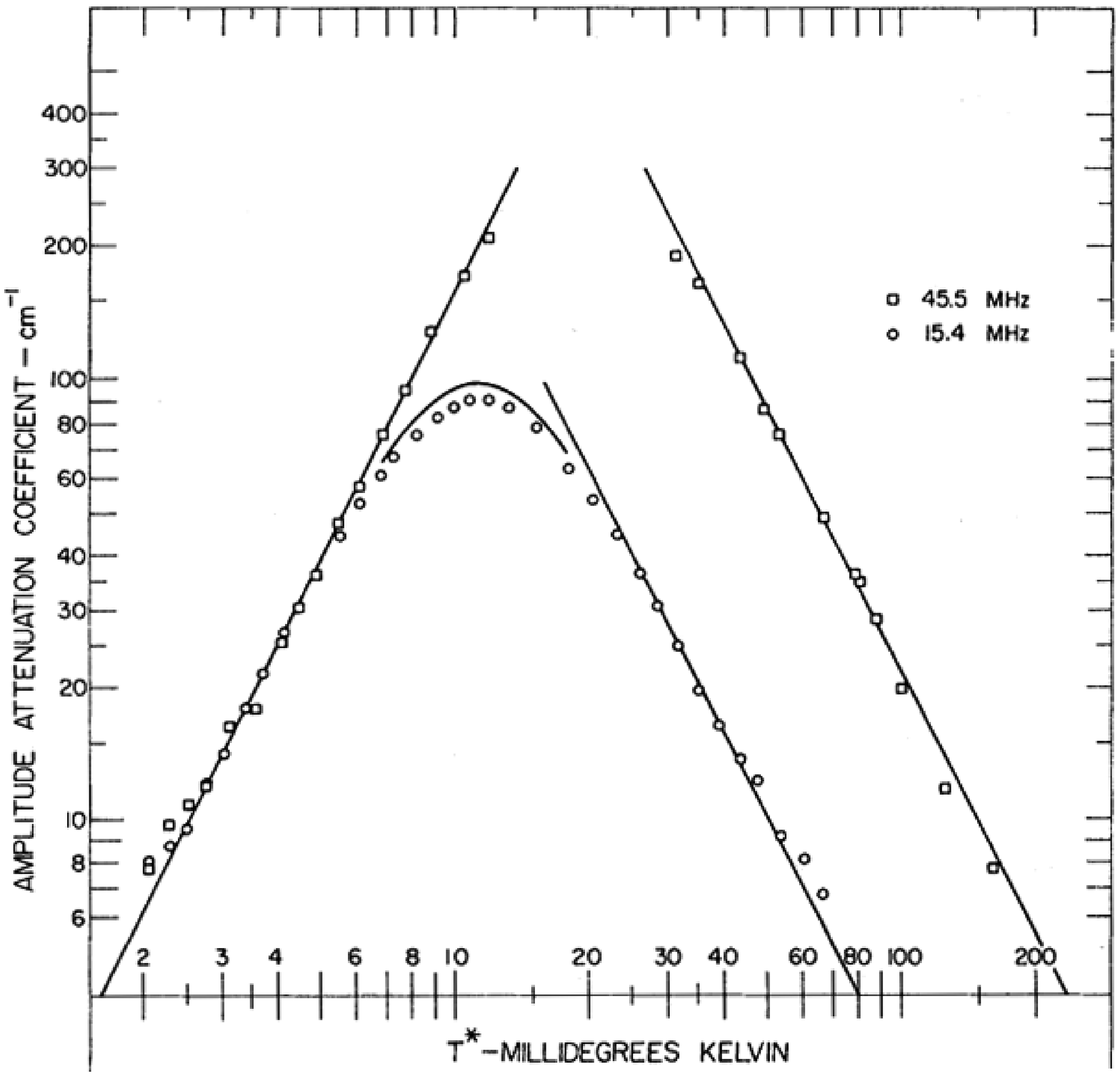}
\caption{ {\small The temperature dependence of the sound attenuation coefficient $\Gamma_q$ in various regimes of a Landau Fermi-liquid. Above:
 A sketch of the dependence in the hydrodynamic (I), 
collisionless thermal (II) and collisionless quantum (III) regimes. Below: Temperature dependence of the acoustic attenuation in liquid $\,^3$He at P=32 kPa measured at both 15.4 
MHz ($\bigcirc$) and 45.5 MHz ($\Box$). The lines through the data correspond to 
$\log \Gamma_q \sim 2 \log T$ and $\log \Gamma_q \sim - 2 \log T$ in the collisionless thermal and hydrodynamic regimes, respectively, in agreement with Table \ref{tab1}. Reprinted with permission from Abel {\it et al.}\, 
 \protect\cite{Abel:1966zz}. Copyright (1966) by the American Physical Society. }}
\label{fig:log-gamma-plots-vertical}
\end{figure*}

The sound attenuation constants 
in various regimes are shown in Table \ref{tab1} and Fig.~\ref{argqplot}. The temperature dependence of the sound attenuation coefficient $\Gamma_q$ is shown in Fig.~\ref{fig:log-gamma-plots-vertical}. We shall use this information as a suggestive 
guide in our investigation of the holographic zero sound at finite temperature.

In the massless D3/D7 system at low temperature, the chemical potential is proportional to the cubic root of the 
volume density $d$ of the $U(1)$ ``baryon" charge\footnote{An explicit expression for the charge density 
operator involving fundamental fermions and complex scalars of the ${\cal N}=2$ hypermultiplet 
is given in Appendix A of \cite{Kobayashi:2006sb}. \label{footnote1} } \cite{Karch:2008fa}
\begin{equation}
\label{chemical-zero}
\mu = \alpha \, d^{1/3} \left( 1 + O \left( \frac{T}{d^{1/3}}\right) \right)\,,
\end{equation}
where $\alpha$ can be expressed using the Euler beta-function, $\alpha = B(1/3,7/6)/2 \approx 1.402$. 
We shall study the D3/D7 theory in the limit $T\ll d^{1/3}$, 
$\omega \ll d^{1/3}$ formally 
corresponding to the applicability regime (\ref{applicability}) of Landau Fermi-liquid theory.
The appropriate dimensionless variables are
\begin{equation}
\label{def-w-q}
\wbar = \frac{\omega}{d^{1/3}}\,, \qquad \qbar = \frac{q}{d^{1/3}}\,, \qquad \dtilde = \frac{d}{(\pi T)^3}\,.
\end{equation}
{\it A priori}, we do not expect to find agreement with the LFL results outlined above since the 
D3/D7 system appears to be microscopically rather different, with no obviously detectable Fermi surface 
or long-lived quasiparticles in its vicinity. Nevertheless, we do find that the behaviour of the 
zero sound mode at finite temperature is qualitatively similar to the one predicted (and observed) in 
a Landau Fermi-liquid:

i) The three regimes (hydrodynamic, collisionless thermal and collisionless quantum) are readily identified
by analysing the behaviour of the lowest quasinormal frequencies and the spectral function of the 
charge density correlator. The hydrodynamic - collisionless thermal transition occurs 
at $\wbar\, \dtilde^{\, 2/3} \sim 1$ and is most spectacularly manifested in the motion of the zero sound poles 
in the complex frequency plane: as the temperature is increased, the two poles 
(corresponding to zero sound propagation with velocities $+v_s$ and $-v_s$ in the field theory) 
recede into the complex plane, approximately tracing a circle, until they collide on the imaginary axis and form two poles with zero real part - one 
of these new poles recedes even deeper into the complex plane while the other approaches the origin. The latter pole is the charge density diffusion mode (with the diffusion constant computed in \cite{Kim:2008bv,Mas:2008qs}) 
characteristic of the hydrodynamic regime. 
Such behaviour was previously observed in \cite{Kaminski:2009dh} 
and also in a (2+1)-dimensional holographic field theory in \cite{Bergman:2011rf}, where it was correctly identified as the hydrodynamic - collisionless transition involving the zero sound mode. 
A similar transition between propagating and diffusive modes has been seen in a model 
of a holographic superconductor \cite{Herzog:2010vz}. The second transition, between the collisionless thermal and the collisionless quantum regimes, is observed at  $\wbar\, \dtilde^{\, 1/3} \sim 1$.

As our investigation is limited  to the current-current correlators, we are not able to follow the emergence of first sound in the hydrodynamic regime - this should appear as a pole in the energy-momentum tensor correlators of the field theory which are decoupled from the current-current correlators in the probe brane limit. Accordingly, we do not expect the details of the 
hydrodynamic - collisionless thermal transition described above to survive beyond the probe brane 
approximation: it seems more likely that the acoustic poles, rather than colliding on the imaginary axis 
as the temperature is raised, will recede deeper into the complex plane, reflecting the behavior of the attenuation curve in Fig.~\ref{fig:log-gamma-plots-vertical}, and then come back close to the real axis again as the hydrodynamic sound poles.

ii) The D3/D7 zero sound attenuation coefficients exhibit the momentum, temperature and density dependence 
typical of a Landau Fermi-liquid as shown in Table \ref{tab1}. We find that the dependence of the acoustic damping upon
frequency and temperature is qualitatively the same as shown in 
Figs.~\ref{argqplot}, \ref{fig:log-gamma-plots-vertical} in regions II and III.

iii) These results remain valid in the case of a non-vanishing hypermultiplet mass.\footnote{Within the region of thermodynamic stability of the theory.}

The structure of the paper is as follows. In Section
\ref{sec:setup}, we give a brief description of the gravitational dual of the D3/D7 
field theory
and the relevant, known properties of its Green's functions and
spectral functions. In Section \ref{sec:lowtemperaturebehaviour} we present our numerical results for
the density-density spectral function (and the dominant pole of the
corresponding Green's function) when $T\ll d^\frac{1}{3}$. 
We identify the three regimes similar to those of a Landau Fermi-liquid, 
and describe in detail the behaviour of the collective modes as the temperature of the system is varied.
We summarize our results and discuss how they may
generalize to other holographic finite density systems in Section
\ref{sec:discussion}. Some details relevant for the case of a massive hypermultiplet 
are relegated to appendices: Appendix \ref{appendixA} contains the action and
equations of motion for the fluctuations, and  Appendix \ref{appendixB} provides a 
derivation of the zero sound attenuation constant at zero temperature.

\section{The D3/D7 system}
\label{sec:setup}

The specific field theory whose elementary excitations at finite temperature and density 
we wish to investigate is
(3+1)-dimensional $\mathcal{N}=4$ $SU(N_c)$ supersymmetric Yang-Mills theory coupled
to $N_f$ $\mathcal{N}=2$ fundamental hypermultiplets with a global $U(N_f)$ flavour
symmetry. It arises as the low-energy theory on the worldvolume of a
set of $N_c$ D3-branes and $N_f$ D7-branes intersecting along
(3+1)-dimensions. Taking $N_c\rightarrow\infty$ with both
$\lambda= g_{YM}^2N_c$ and $N_f/N_c$ fixed, and subsequently
taking $\lambda\rightarrow\infty$ and $N_f/N_c\rightarrow0$, we obtain
a classical gravitational dual to this field theory \cite{Karch:2002sh,Kruczenski:2003be}:
\begin{equation}
\begin{aligned} S&=S_{\text{adjoint}}+S_{\text{fundamental}},\\
&=S_{\text{adjoint}}-N_fT_{D7}\int
d^8\xi\sqrt{-\text{det}\left(g_{ab}+F_{ab}\right)},
\end{aligned}
\label{d3d7action}
\end{equation} where $S_{\text{adjoint}}$ is the ten-dimensional
supergravity action and $T_{D7}$ is the tension of a D7-brane. In this
probe brane limit, the metric is fixed and it is $S_{\text{fundamental}}$
which contains the dynamical information. For a zero temperature field
theory, the contribution of the fundamental matter is the DBI action
of $N_f$ probe D7-branes extended along an $AdS_5\times S^3$ section
of the (fixed) $AdS_5\times S^5$ background spacetime generated by the
D3-branes. In Eq.~(\ref{d3d7action}), $g_{ab}$ denotes the induced worldvolume metric 
on the D7-brane and $F_{ab}$ is the field strength of a worldvolume $U(1)\subset U(N_f)$ gauge field.

When the field theory is at a non-zero temperature,
the background spacetime is that of an AdS-Schwarzschild black brane
(with a horizon at $r=r_H$) times a five-sphere:
\begin{equation}
\label{eq:metric}
ds_{10}^2=\frac{r^2}{R^2}\left[ - \left( 1-\frac{r_H^4}{r^4}\right) dt^2+ d \vec{x}^2 \right] + 
\left( 1-\frac{r_H^4}{r^4}\right)^{-1} \frac{R^2}{r^2} dr^2+R^2 ds_{S^5}^2\,.
\end{equation} 
The D7-brane
wraps an asymptotically $AdS_5\times S^3$ section of the metric, with the
horizon radius of the background related to the temperature of
the field theory via $T = r_H/\pi R^2$.

In terms of the dimensionless  
radial coordinate $u =r_H^2/r^2$, the metric
(\ref{eq:metric}) can be written as
\begin{equation}
ds_{10}^2 = \frac{(\pi T R)^2}{u}\left( - f dt^2 + d \vec{x}^2 \right) + \frac{R^2}{4 u^2 f} du^2
+R^2\left(d\theta^2+\sin^2\theta ds_{S^1}^2+\cos^2\theta ds_{S^3}^2\right),
\end{equation} 
where $f(u)=1-u^2$. In these coordinates, the horizon is located at $u=1$ and the
boundary at $u=0$. In equilibrium, the D7-brane embedding can be
characterised by a single embedding coordinate $\theta(u)$ (which
determines which $S^3$ section of the background $S^5$  it
wraps). The gauge field on the brane is dual to a global flavor
current in the field theory and thus turning on the time component of
a $U(1)\subset U(N_f)$ gauge field $A_t(u)$ on the brane corresponds
to introducing a finite density $d$ of the $U(1)$ ``baryon" charge in the field
theory. In this case, it corresponds to a net density of fundamental fermions and scalars.

The equations of motion for the background fields are
obtained from the DBI action
\begin{equation}
\label{eq:action} 
S_{\text{fundamental}}= - \frac{Nr_H^4}{2}\int^1_0 du d^4x\frac{\cos^3\theta}{u^3}\sqrt{1+4u^2f\theta'^2-4\frac{u^3}{r_H^2}A_t'^2}\,,
\end{equation}
 where $N=N_fT_{D7}V_{S^3}$ is a normalization constant determined by 
the gauge-gravity duality dictionary \cite{Kobayashi:2006sb} and primes denote 
derivatives with respect to $u$. One of the equations of motion derived from the action (\ref{eq:action}) reduces to 
\begin{equation}
A_t'(u)=-\frac{r_H\dtilde}{2}\sqrt{\frac{1+4u^2f(u)\theta'(u)^2}{\cos^6\theta(u)+\dtilde^2u^3}}\,,
\end{equation}
where $\dtilde\equiv d R^6/r_H^3 = d/\left(\pi T\right)^3$
is a dimensionless parameter of the field theory related to the 
net number density of ``quarks" $n_q$ in the
field theory\footnote[2]{See footnote \ref{footnote1}. Note that our normalization of $d$ is different from the one used in \cite{Kobayashi:2006sb}: in 
\cite{Kobayashi:2006sb}, $d\sim n_q$ whereas in our case $d\sim n_q/\sqrt{\lambda} N_c N_f$.} via
\begin{equation}
\dtilde=\frac{2^{5/2}n_q}{\sqrt{\lambda} N_f N_cT^3} \,.
\end{equation}  
For a given density $\dtilde$, there is a corresponding dimensionless chemical
potential given by
\begin{equation}
\label{eq:muintegral}
\mutilde=\frac{\dtilde}{2}\int_0^1du\sqrt{\frac{1+4f(u)u^2\theta'(u)^2}{\cos^6\theta(u)+\dtilde^2u^3}}\,,
\end{equation} 
which is related to the field theory chemical potential $\mu_{FT}$ by $\tilde{\mu}\equiv\mu/\pi T=\sqrt{\frac{2}{\lambda}}\frac{\mu_{FT}}{T}$. 
In the massless case, the integral in Eq.~(\ref{eq:muintegral}) can be computed 
exactly \cite{Karch:2008fa}: it reduces to (\ref{chemical-zero}) in the low temperature limit. 
The important scaling (not entirely obvious from  Eq.~(\ref{eq:muintegral}) and generically 
accompanied by mass and temperature corrections \cite{Karch:2008fa,Karch:2007br}) to emphasize is
\begin{equation}
\label{mu-scales}
 \frac{\mu}{T} \sim \mutilde \sim \dtilde^{1/3}\,.
\end{equation}

The equation of motion for the embedding coordinate is
\begin{equation}
\frac{d}{du}\left(\frac{f(u)\cos^3\theta(u)\theta'(u)}{u\sqrt{G(u)}}\right)
+\frac{3\cos^2\theta(u)\sin\theta(u)\sqrt{G(u)}}{4u^3}=0\,,
\label{emb-eq}
\end{equation} 
where
$$
G(u) = \cos^6\theta(u)\frac{1+4u^2f(u)\theta'(u)^2}{\cos^6\theta(u)+u^3\dtilde^2}\,.
$$
Eq.~(\ref{emb-eq}) has no known generic analytic solution and must be solved
numerically\footnote{A method to approximate the solution in the low
temperature limit has been proposed in \cite{Wapler:2010nq}.}. 
Near the boundary, the solution has the form
\begin{equation} \theta(u)=\frac{\mtilde}{\sqrt{2}}\sqrt{u}+\ldots,
\end{equation} 
where $\mtilde$ is related to the bare mass\footnote{The explicit field theory  ``quark" mass
operator is given in Appendix A  of \cite{Kobayashi:2006sb}.}
 of the
fundamental matter in units of temperature via \cite{Mateos:2007vn}
\begin{equation}
\label{eq:massdefinition}
\mtilde=\frac{2}{\sqrt{\lambda}}\frac{M_q}{T}.
\end{equation} 
To study the temperature dependence of the theory at finite density, it will be more 
convenient to use the field theory mass normalized with respect to density 
rather than temperature:
\begin{equation}
\mbar=\frac{\mtilde}{\dtilde^{\frac{1}{3}}}=
\frac{2\pi}{\sqrt{\lambda}}\frac{M_q}{d^{\frac{1}{3}}}\,.
\end{equation}

Note that $\theta(u)=0$ for a massless hypermultiplet, and
thus the dual description greatly simplifies in this case - there is
only one non-trivial background field $A_t(u)$ which has the simple
equation of motion
\begin{equation}
A_t'(u)=-\frac{r_H\dtilde}{2}\left( 1+\dtilde^2u^3\right)^{-1/2}\,.
\end{equation}

The thermodynamics of the system can be determined by computing the on-shell action. In the 
massless case, the specific heat
schematically has the following temperature dependence \cite{Karch:2008fa}:
\begin{equation}
c_V \sim N^2_c T^3 + \cdots + \lambda N_f N_c T^6/d +\cdots\,,
\end{equation} 
where the first term comes from the adjoint ${\cal N}=4$ SYM degrees of freedom. The pressure $P$, the 
entropy density $s$ and the energy density $\varepsilon$ can be computed from the 
thermodynamic potential $\Omega (\mu,T)$ given explicitly in \cite{Karch:2008fa}: the entropy density 
is finite in the $T\rightarrow 0$ limit, 
$s\sim \mu^3$, and the equation of state is $\varepsilon = 3 P$ for all $T$ and $\mu$ 
(and thus the speed of first sound is $v_s=1/\sqrt{3}$). 
In certain regimes, the system becomes thermodynamically unstable: there appears to be 
 an instability at low density
$\dtilde < \dtilde_c \approx 0.00315$ \cite{Kobayashi:2006sb,Kaminski:2009ce}, and at high density and large mass
$\mbar\gg1, \dtilde\gg1$ \cite{Wapler:2010nq}. In addition, when $d\ne0$ and $m\ne0$ there is a lower limit on the possible value of $\mu$, below which there is a phase transition to a Minkowski embedding with $d=0$
\cite{Nakamura:2007nx,Ghoroku:2007re,Karch:2007br,Faulkner:2008hm,Mateos:2007vc}. The values of the parameters considered in this paper are 
outside of these regimes.

To determine the field theory excitations, one considers fluctuations of the background bulk fields
\begin{equation}
\begin{aligned} &\theta(u)\rightarrow\theta(u)+\phi(u,z,t)\,,\\
&A_t(u)\rightarrow A_t(u)+a_t(u,z,t)\,,\\ &A_i(u)\rightarrow a_i(u,z,t)\,,
\end{aligned}
\end{equation} 
$i=x,y,z$, where we have chosen the fluctuations to depend on time, the radial coordinate 
and one of the spatial coordinates ($z$) only (the latter is possible due to the isotropy of the theory).
Introducing Fourier components for the embedding fluctuation
\begin{equation} 
\phi(u,z,t)=\int\frac{d\omega \, dq}{\left(2\pi\right)^2}e^{-i \omega t + i qz} \, \tilde{\phi} \left(u,\omega,q\right),
\end{equation} 
and similarly for $a_\mu(u,z,t)$, and expanding the DBI action (\ref{eq:action}) to quadratic
order in fluctuations, one finds the resulting action and the corresponding
equations of motion. The transverse modes
$\tilde{a}_x$, $\tilde{a}_y$ decouple from the longitudinal modes $\tilde{a}_t$,
$\tilde{a}_z$ and $\tilde{\phi}$. In the following, we shall focus on the longitudinal modes
as they are the ones that encode the fate of zero sound at finite temperature.

For a non-zero hypermultiplet mass, the equations of motion 
lead to a pair of coupled
differential equations for the embedding fluctuation
$\tilde{\phi}\left(u,\wbar, \qbar \right)$ and the gauge-invariant
combination (c.f. \cite{Kovtun:2005ev})
\begin{equation}
\label{eq:gicombo}
\Zbar(u,\wbar,\qbar) = \wbar\, \tilde{a}_z(u,\wbar,\qbar) + \qbar \, \tilde{a}_t(u,\wbar,\qbar)\,,
\end{equation} 
where the dimensionless variables $\wbar$ and $\qbar$ were introduced in Eq.~(\ref{def-w-q}). 
The equations of motion and the action for the longitudinal fluctuations are given
 in Appendix \ref{appendixA}.

In the zero mass limit, the equations for these two modes
decouple and the gauge-invariant combination (\ref{eq:gicombo}) obeys
the equation of motion
\begin{equation}
\label{eq:masslesseom}
\frac{d}{du}\left[ \frac{f(u)\,  \Zbar'}{\sqrt{g(u)}\left(\wbar^2-\qbar^2 f g \right)}\right] 
+ \frac{\dtilde^{\frac{2}{3}} \bar{Z}}{4 u \sqrt{g(u)} f(u)}=0,
\end{equation} 
where $g(u)=(1+\dtilde^2u^3)^{-1}$.
In this limit, the longitudinal part of the on-shell
action is
\begin{equation} S^{(2)}_{\text{long.}}=Nr_H^2 \int \frac{d\omega
dq}{\left(2\pi\right)^2} \frac{f\, \Zbar(u,-\wbar,-\qbar)\Zbar'(u,\wbar,\qbar)}{\sqrt{g}
\left(\wbar^2-\qbar^2 f g \right)}\Biggr|^{u_B}_{u_H}.
\end{equation}

In this massless case, we obtain the longitudinal retarded
Green's functions from the gravitational fields via the usual procedure
\cite{Son:2002sd}
\begin{equation}
\label{eq:longitudinalgreens}
G^R_{J^zJ^z}\left(\wbar,\qbar\right) 
= -\lim_{\epsilon\rightarrow 0}  2 Nr_H^2 \frac{\wbar^2}{\wbar^2-\qbar^2}\, 
\frac{\Zbar'(\epsilon,\wbar,\qbar)}{\Zbar(\epsilon,\wbar,\qbar)}\,,
\end{equation}
where the $-\wbar^2$ factor comes from the
definition (\ref{eq:gicombo}), and
$\Zbar(u,\wbar,\qbar)$ is the solution obeying ingoing
boundary conditions at the horizon:  $\Zbar(u,\wbar,\qbar)\sim\left(1-u\right)^\gamma$ as $u\rightarrow1$, where 
$\gamma  = - i \wbar \dtilde^{1/3}/4$.
Poles of the retarded Green's function are determined by the values of
$\wbar\left(\qbar\right)$ for which the solution obeying the ingoing condition 
at the horizon vanishes at the boundary \cite{Kovtun:2005ev,Son:2002sd}. The density-density correlation function follows trivially
from the conservation of current $G^R_{J^tJ^t}\left(\wbar,\qbar\right) = \qbar^2\, G^R_{J^zJ^z}/\wbar^2$ and has the same poles as the longitudinal Green's
function (\ref{eq:longitudinalgreens}). The spectral functions of
these operators are then obtained by taking the imaginary part: 
\begin{equation}
\chi_{zz}\left(\wbar, \qbar\right) = - 2\, \text{Im}\left[G^R_{J^zJ^z}\left(\wbar,\qbar\right)\right]\,, 
\qquad 
\chi_{tt}\left(\wbar, \qbar\right) = - 2\, \text{Im}\left[G^R_{J^tJ^t}\left(\wbar,\qbar\right)\right]\,, 
\end{equation} 
with $\chi_{tt}=\qbar^2\, \chi_{zz}/\wbar^2$. At zero temperature and density, the form of the
longitudinal spectral function is known analytically \cite{Myers:2007we}: 
\begin{equation}
\label{eq:spectralasymptotics}
\chi_{zz}\left(\omega,q\right) =\frac{N_f N_c}{4\pi}\, \left(\omega^2-q^2\right)\, \Theta \left(\omega^2-q^2\right)
\text{sgn}\, \omega\,,
\end{equation} 
where $\Theta$ is the Heaviside step function.

In the $\mbar\ne0$ case, the coupled equations
of motion for bulk fluctuations imply that the dual field theory operators mix and the method
to determine the retarded Green's functions is more involved (see Appendix \ref{appendixA} for the details).

At  zero mass, high densities $\omega, q\ll
d^{\frac{1}{3}}$ and strictly zero temperature, the dominant pole of the correlators $G^R_{J^tJ^t}$ and 
$G^R_{J^zJ^z}$ has the dispersion relation \cite{Karch:2008fa,Nickel:2010pr}
\begin{equation}
\label{eq:zerosound}
\wbar=\pm\frac{\qbar}{\sqrt{3}}-i\frac{\Gamma\left(\frac{1}{2}\right)}{\Gamma\left(\frac{1}{6}\right)\Gamma\left(\frac{1}{3}\right)}\qbar^2+O\left(\qbar^3\right).
\end{equation} This corresponds to a collective excitation of the
system - the holographic zero sound mode. Its speed is equal to that of
hydrodynamic sound, and it has an imaginary part $\propto q^2$. 
In the hydrodynamic limit $\omega, q\ll  T$ the
dominant pole is a purely imaginary pole with the Fickian diffusion dispersion relation
\begin{equation}
\label{eq:diffusion}
\wbar = - i D \, \qbar^2 + O(\qbar^3)\,,
\end{equation} 
where the diffusion constant is given by \cite{Kim:2008bv,Mas:2008qs}
\begin{equation} 
D(\dtilde)=\frac{\dtilde^{\frac{1}{3}}}{2}\sqrt{1+\dtilde^2}\;_2F_1\left[\frac{3}{2},\frac{1}{3};\frac{4}{3};-\dtilde^2\right].
\end{equation} 
Between these two extreme temperature limits, the poles
of the Green's functions are not known analytically.

These results were generalized to the case of a massive hypermultiplet in \cite{Kulaxizi:2008kv,Mas:2008qs}. The zero sound mode (\ref{eq:zerosound}) persists when the hypermultiplet has 
a finite mass $m$, although the dispersion relation is altered to 
\begin{equation}
\label{eq:massivezerosound}
\wbar = \pm  \frac{1}{\sqrt{3}}\,  \Biggl( \frac{1- \mgot^2}{1-\mgot^2/3} \Biggr)^{1/2}   \qbar  \,  - \,    
i \, \frac{\Gamma\left(\frac{1}{2}\right)}{\Gamma\left(\frac{1}{3}\right)\Gamma\left(\frac{1}{6}\right)}\, 
\frac{(1-\mgot^2)^{4/3}}{ (1-\mgot^2/3)^2}   \,  \,  \qbar^2    +   O\left(\qbar^3\right),
\end{equation}
where 
$\mgot=\tilde{m}/\sqrt{2}\tilde{\mu}=M_q/\mu$.
The real part of the dispersion relation
 (\ref{eq:massivezerosound}) was obtained in \cite{Kulaxizi:2008kv} and the attenuation is derived in Appendix \ref{appendixB}. In the hydrodynamic limit, there is again a diffusion pole (\ref{eq:diffusion}) whose diffusion constant can be derived via an Einstein relation \cite{Mas:2008qs}
 \begin{equation}
 \begin{aligned}
 D(\dtilde)=\frac{\dtilde^{\frac{1}{3}}}{2}\sqrt{\dtilde^2+\cos^6\theta(1)}\int^1_0du&\frac{G(u)^\frac{3}{2}}{\cos^3\theta(u)H(u)}\biggl[1+\dtilde\biggl(\frac{4f(u)u^2\theta'(u)}{G(u)\cos\theta(u)}\frac{\partial}{\partial\dtilde}\left[\cos\theta(u)\theta'(u)\right]\\
 &+\frac{\sin\theta(u)}{\cos^2\theta(u)}\left(3+\frac{4f(u)u^2\theta'(u)^2}{G(u)}\right)\frac{\partial}{\partial\dtilde}\sin\theta(u)\biggr)\biggr],
 \end{aligned}
 \end{equation}
 where $G(u)$ and $H(u)$ are defined in Appendix \ref{appendixA}.

At finite hypermultiplet mass, the D3/D7 system has non-trivial bound states analogous to mesons 
\cite{Kruczenski:2003be}. These bound states are visible as peaks of the spectral function. When $d=0$, such modes exist only for large enough values of $\mtilde\propto M_q/T$ 
\cite{Mateos:2006nu,Mateos:2007vn,Hoyos:2006gb,Myers:2007we,Paredes:2008nf}. When $d\ne0$, it seems that necessary conditions for their existence are
large values of $\mtilde$ and small enough values of $q/T$ and
$d^{\frac{1}{3}}/T$ \cite{Erdmenger:2007ja,Myers:2008cj,Mas:2008jz,Kaminski:2009ce,Kaminski:2009dh}. This is outside of the regime of our current interest. A full numerical analysis of the quasinormal modes of the theory when $T=0$ is given in \cite{Ammon:2011hz}.

\section{The high density regimes of D3/D7 fundamental matter}
\label{sec:lowtemperaturebehaviour}

In this section, we explore the behavior of the collective modes of the D3/D7 system in the low 
temperature regime $\dtilde \gg 1$ ($T \ll \mu$). Anticipating behavior similar to that observed in a Landau Fermi-liquid, we expect the system to exhibit 
the three regimes shown in Fig.~\ref{arrow-scales}, and indeed we 
do find them. We investigate this by computing numerically the spectral functions 
and the poles of the Green's functions using the approach of 
\cite{Son:2002sd,Kaminski:2009dh}. This approach provides a numerical consistency 
check (invariance under radial translations) which we used to ensure our results were accurate.
We always plot the normalized spectral functions such as
$\chi_{zz}\left(\wbar,\qbar\right)/2N_fN_cT^2\wbar^2$: dividing by $\wbar^2$ has the advantage of 
reducing the high frequency asymptotics (\ref{eq:spectralasymptotics}) to a
constant.

\subsection{The collisionless quantum regime}
\label{sec:cqr}

\begin{figure}
\begin{center}
\includegraphics[scale=0.83]{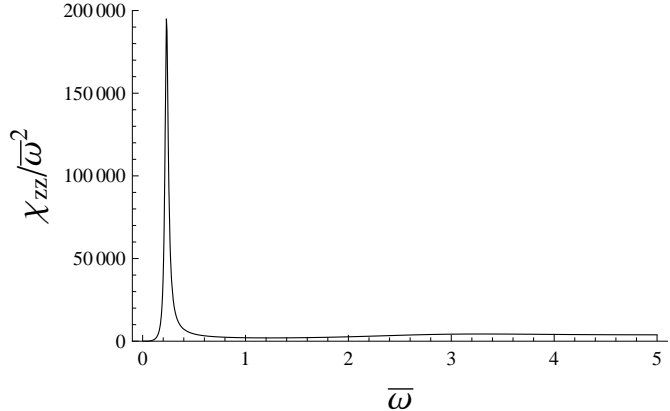}
\caption{The holographic zero sound peak in the collisionless quantum regime. 
The longitudinal spectral function is shown at $\mbar=0$,
$\dtilde=10^6$, $\qbar=0.4$.}
\label{fig:zerosoundspectral}
\end{center}
\end{figure}

The collisionless quantum regime corresponds to $\wbar$ and $\qbar$ being in the interval
\begin{equation}
\label{interval-cqr}
\dtilde^{-1/3} \ll \wbar, \qbar \ll 1\,,
\end{equation}
see Eq.~(\ref{mu-scales}), Fig.~\ref{arrow-scales} and Table \ref{tab0}. 
We start by fixing a large value of $\dtilde$, 
for example $\dtilde=10^6$,
and computing the longitudinal spectral function at a fixed momentum $\qbar$ in 
the interval (\ref{interval-cqr}), e.g. $\qbar =0.4$. An isolated peak corresponding to the 
holographic zero sound mode in this low temperature 
regime is clearly visible in Fig.~\ref{fig:zerosoundspectral}. 
Thus the zero sound mode persists at low but non-zero temperatures - contrary 
to the assertion in \cite{Kim:2008bv} that an infinitesimal temperature in the
field theory will lead to diffusive transport.

By varying $\qbar$ (while keeping $\dtilde^{\frac{1}{3}}$ fixed), we can determine the dispersion
relation of this collective mode at low temperature. 
Figure \ref{fig:masslessdispersion} shows the real and imaginary parts of the zero sound 
dispersion relation $\wbar = \wbar(\qbar)$ at $\mbar =0$, $\dtilde=10^6$. Our numerical results are 
shown as dots and the solid lines correspond to the zero-temperature dispersion relation (\ref{eq:zerosound}).
We observe that in the collisionless quantum regime  (\ref{interval-cqr}) 
the real part of the zero sound dispersion relation shows no noticeable deviation from the 
zero-temperature result (\ref{eq:zerosound}). The imaginary part shows a close
agreement with the zero-temperature result over the parameter range  (\ref{interval-cqr}). 
Visible deviations from the zero-temperature result are apparent at very small
($\qbar\lesssim0.02 \sim \dtilde^{-1/3}$) and very large ($\qbar\gtrsim0.7$) momenta 
close to the boundaries of the interval (\ref{interval-cqr}). Note that
although the imaginary part appears to be tending to a constant at small $\qbar$, this
is only true up until the crossover to the hydrodynamic regime occurs,
after which we obtain a diffusive mode. This crossover will be discussed in detail
later in this section. At large momentum, the system is entering the low
temperature, low density regime $\dtilde^{-1/3} \ll 1 \ll \wbar, \qbar$ 
(i.e. $\omega, q \gg d^{\frac{1}{3}}\gg T$), where the zero sound 
mode becomes very short-lived as the corresponding pole recedes deep into the complex plane. 
As shown in Fig.~{\ref{fig:masslessasymptotics}, 
the zero sound peak in the spectral function gradually disappears in this regime, and 
the spectral function approaches the $d=0$, $T=0$ result (\ref{eq:spectralasymptotics}). 
\begin{figure*}
\begin{center}
\includegraphics[scale=0.85]{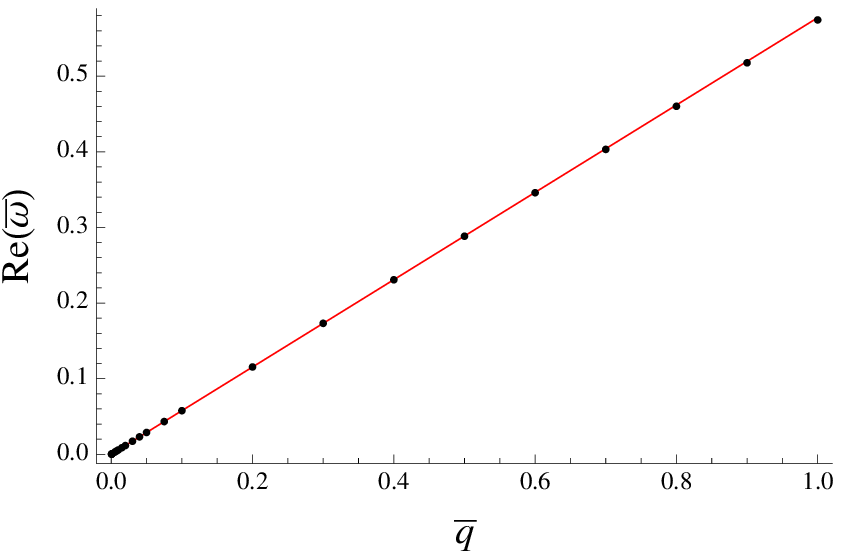}
\includegraphics[scale=0.85]{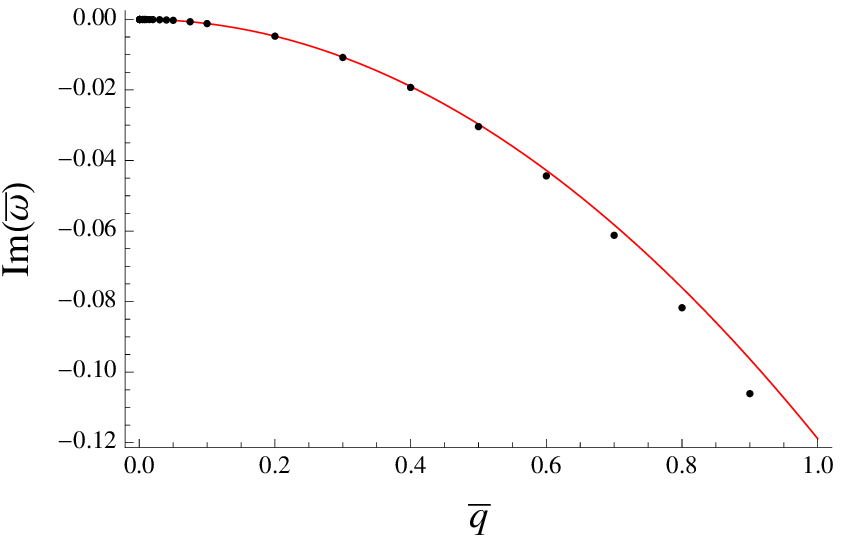}
\includegraphics[scale=0.85]{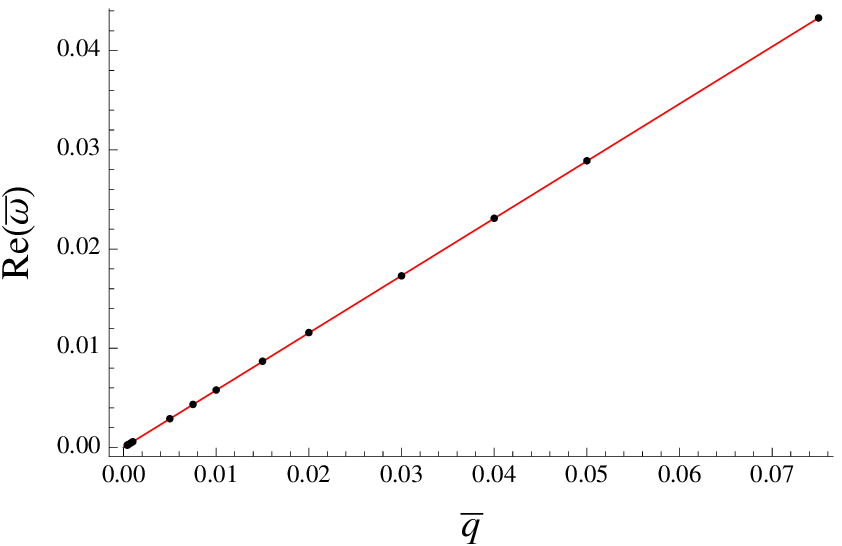}
\includegraphics[scale=0.85]{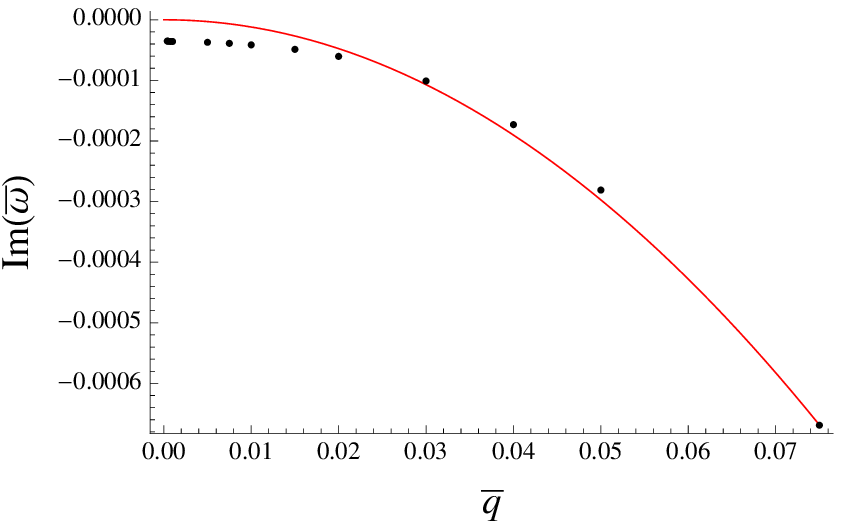}
\caption{The dispersion relation of the dominant pole at
$\mbar=0$, $\dtilde=10^6$. Dots show numerical results
at low $T$ and the solid lines are the analytic result
(\ref{eq:zerosound}) at $T=0$.}
\label{fig:masslessdispersion}
\end{center}
\end{figure*} 

At finite hypermultiplet mass, the results are qualitatively similar.\footnote{Technically, 
the massive case is more complicated as it involves solving a pair of coupled 
differential equations, with coefficients that depend upon the numerically-computed embedding function. Due to numerical instability, we were not able to obtain 
accurate results for $\dtilde>10^5$ and masses ouside the interval 
$0.002 \lesssim \mbar \lesssim1.68$ (at $\dtilde=10^5$). 
We believe this limitation demonstrates the lack of our numerical skills 
rather than an effect of any physical significance.} The longitudinal spectral functions 
show a lone peak\footnote{Peaks corresponding to ``meson" bound states may 
exist for higher values of $\wbar$. We have not investigated this issue.}
(similar to the one shown in Fig.~\ref{fig:zerosoundspectral}) for values of $\wbar$ in the range (\ref{interval-cqr}). 
As shown in Fig.~\ref{fig:massivedispersion1}, the real part of the dispersion relation 
is essentially identical to the zero-temperature 
result (\ref{eq:massivezerosound}),
and the imaginary part deviates from the zero-temperature dependence given in 
(\ref{eq:massivezerosound}) only
at the boundaries of the interval (\ref{interval-cqr}).  
\begin{figure*}
\begin{center}
\includegraphics[scale=0.7]{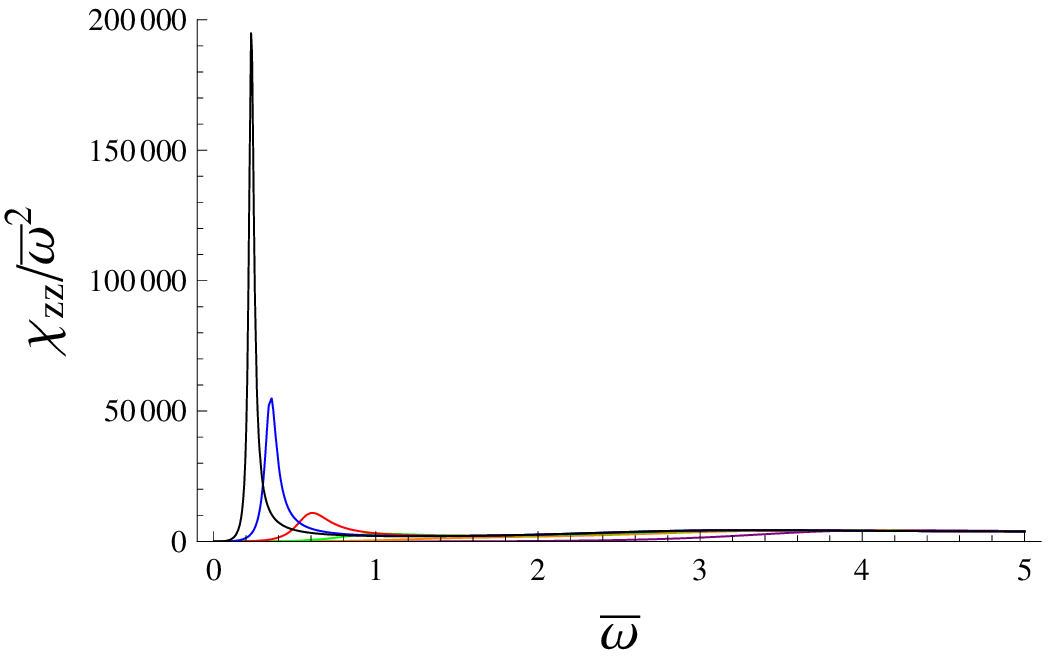}
\includegraphics[scale=0.7]{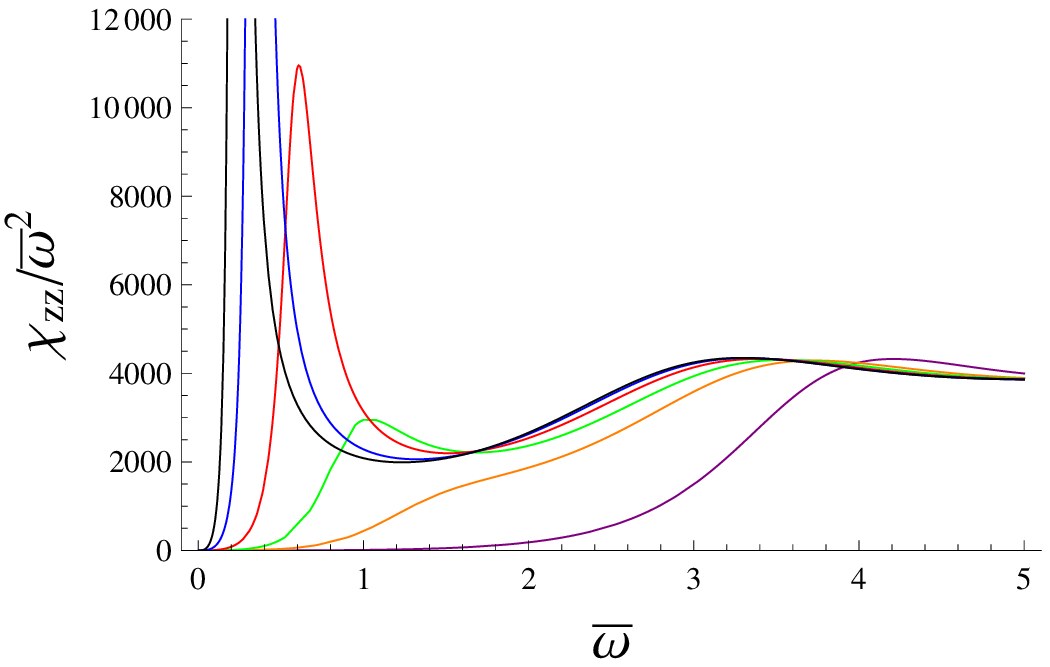}
\caption{The longitudinal spectral function at $\mbar=0$,
$\dtilde=10^6$. Moving from left to right corresponds to increasing
momentum: $\qbar=0.4$ (black), 0.6 (blue), 1.0 (red), 1.5
(green), 2.0 (orange), 3.0 (purple).}
\label{fig:masslessasymptotics}
\end{center}
\end{figure*}

It is interesting to note that the zero sound
mode exists for all (numerically accessible) values of $\mbar$,
including those for which $\mtilde\propto M_q/T\ll1$. Thus the high
density and low temperature interval in which the $T=0$ zero sound mode
persists is the same as in the massless case (given by the inequality 
(\ref{interval-cqr})), i.e. we do not have to take the limit 
$M_q/T\rightarrow\infty$ which would be the
genuine zero temperature limit of the entire system.
\begin{figure*}
\begin{center}
\includegraphics[scale=0.85]{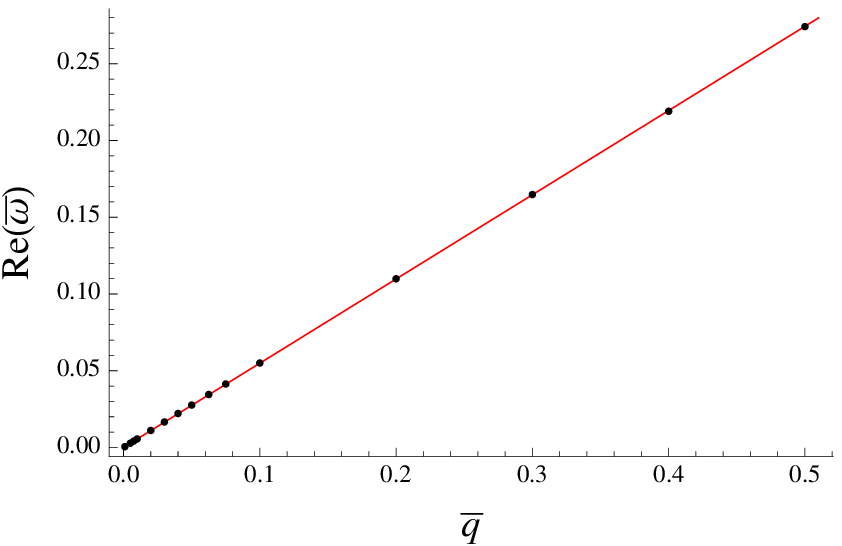}
\includegraphics[scale=0.85]{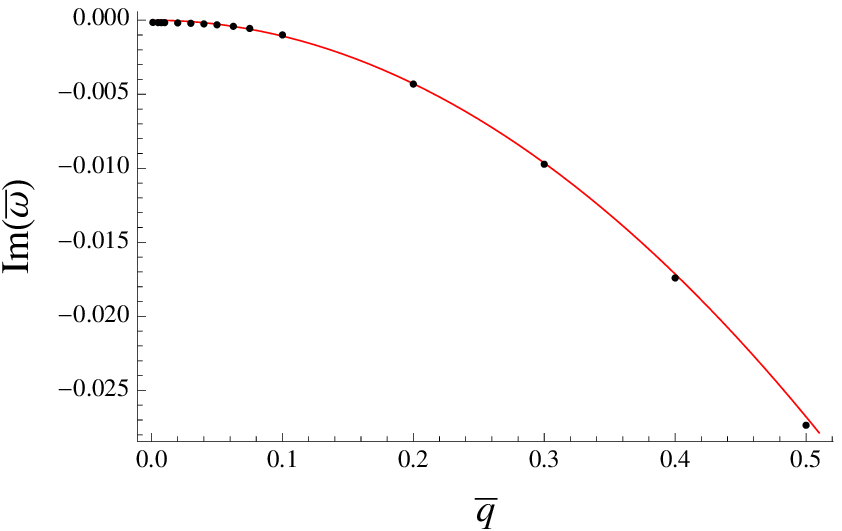}
\includegraphics[scale=0.85]{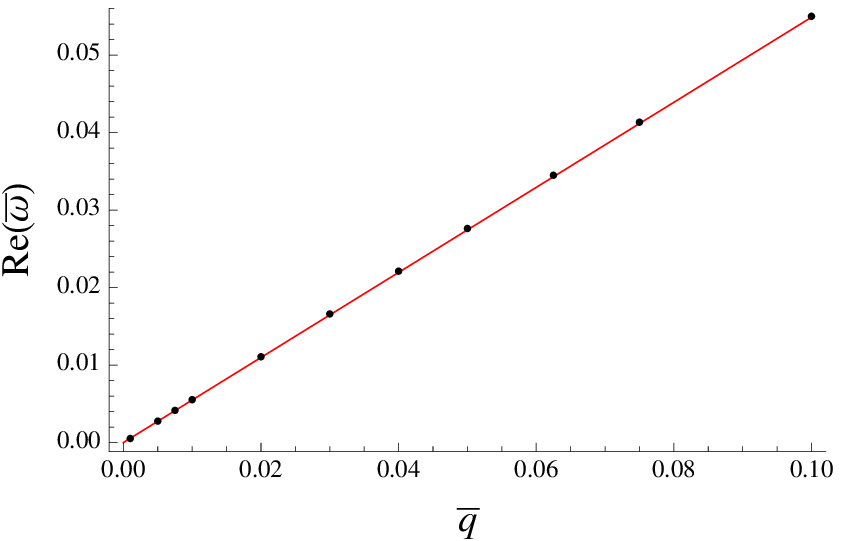}
\includegraphics[scale=0.85]{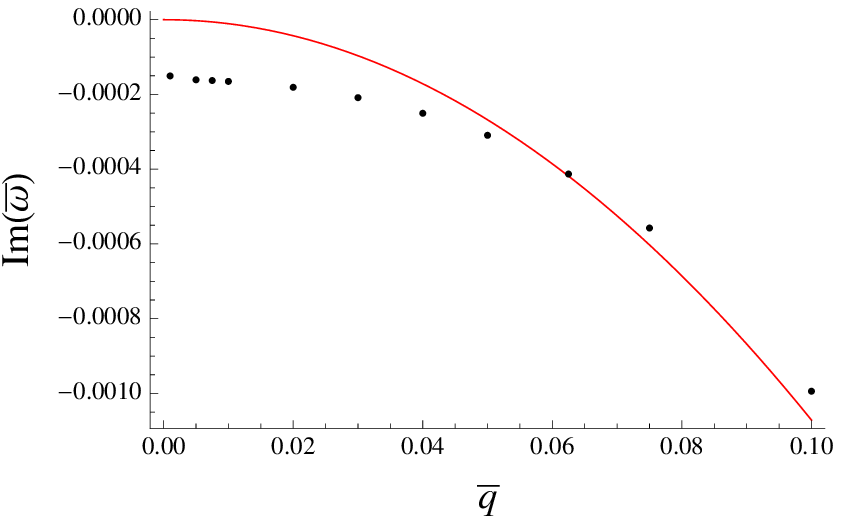}
\caption{The holographic zero sound dispersion relation for $\mbar=0.76$, $\dtilde=10^5$. 
Dots show numerical results at low $T$ and the solid lines show the analytic result
(\ref{eq:massivezerosound}) at $T=0$.}
\label{fig:massivedispersion1}
\end{center}
\end{figure*}
\begin{figure*}
\begin{center}
\includegraphics[scale=0.75]{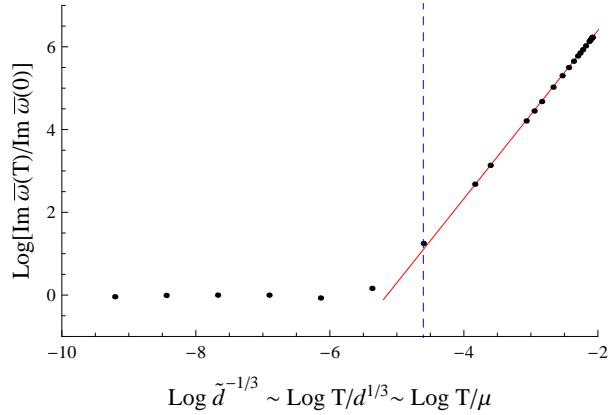}
\caption{The imaginary part of the holographic zero sound dispersion relation
 at low temperatures showing the transition between the collisionless quantum (left) 
and collisionless thermal (right) regimes. The points are the numerical data, the dashed line 
denotes $\qbar=\dtilde^{-1/3} \sim T/\mu$ 
and the solid line is the best-fit straight line for the $T/\mu \gtrsim \qbar = 0.01$ points showing 
the $\sim T^2$ scaling of the attenuation in the collisionless thermal regime. 
The region corresponding to the hydrodynamic regime at even higher temperature 
$T/\mu \gg \sqrt{\qbar} = 0.1$ is not shown in the figure.}
\label{fig:lowTcorrections}
\end{center}
\end{figure*} 
\subsection{The collisionless thermal regime}
\label{sec:ctr}

Having established the existence of a holographic zero sound mode at finite temperature, 
we now investigate the dependence of its velocity and attenuation on temperature, density, momentum and mass.
The collisionless thermal regime corresponds to excitations with $\wbar$ and $\qbar$ in the interval
\begin{equation}
\label{interval-ctr}
\dtilde^{-2/3} \ll \wbar, \qbar \ll \dtilde^{-1/3} \,.
\end{equation}
As discussed in Section \ref{sec:cqr}, the imaginary part of the dispersion relation $\wbar = \wbar (\qbar)$ for 
the zero sound shows significant deviations from the zero temperature $\sim \qbar^2$ behavior in the 
region $\qbar \lesssim \dtilde^{-1/3}$. Investigating this further, one can see that in the interval 
$\dtilde^{-2/3} \lesssim \qbar \lesssim  \dtilde^{-1/3}$ the imaginary part $\mbox{Im}\, \wbar$ is essentially 
independent of $\qbar$. This is the characteristic behavior of the sound attenuation coefficient of a 
Landau Fermi-liquid in the collisionless thermal regime (see Table \ref{tab1}). 
To determine the temperature dependence of 
the attenuation in this regime, in Fig.~\ref{fig:lowTcorrections} 
we plot the logarithm of the ratio $\mbox{Im}\, \wbar (T)/\mbox{Im}\, \wbar (0)$ 
versus the logarithm of $\dtilde^{-1/3} \sim T/\mu$ at fixed $\qbar = 0.01$. In the collisionless quantum regime, 
the attenuation is essentially 
temperature-independent, whereas in the collisionless thermal regime it scales as $\sim T^2$. 
The transition occurs at $\qbar \sim \dtilde^{-1/3}$. This is fully compatible with LFL behavior (see Table \ref{tab1} and Fig.~\ref{fig:log-gamma-plots-vertical}). Fig.~\ref{fig:lowTcorrections} 
should be compared to the transition between regions III and II in 
Fig.~\ref{fig:log-gamma-plots-vertical}.

At non-zero hypermultiplet mass the results are qualitatively the same and the 
plots for $\mbar = 0.76, 1.68$ are similar to Fig.~\ref{fig:lowTcorrections}. We do not show these 
plots for 
conciseness.

Another comparison with Landau Fermi-liquid theory can be made by plotting the attenuation as a 
function of (real) frequency. As shown in Fig.~\ref{argqplot} and Table \ref{tab1}, in 
the frequency dependence of the acoustic attenuation, the transition 
from the collisionless thermal regime to the collisionless quantum regime is characterized by a change 
of scaling from $\text{arg}(q)\propto1/\omega$ to $\text{arg}(q)\propto\omega$ at 
$\wbar \sim T/\mu$. We can study this region in the D3/D7 system by fixing the temperature (at e.g. $\tilde{d}=10^4$) 
and tracing the Green's function's pole in the complex momentum plane while varying the real frequency $\wbar$.
The results are shown in Fig.~\ref{fig:momentumdependence} for the massless case. In the collisionless 
quantum regime, the points follow a straight line with gradient 1. In the collisionless thermal regime, 
for a region where there is a power law dependence $\text{arg}(\bar{q})\propto\bar{\omega}^\alpha$, the 
best fit value of $\alpha$ turns out to be $\alpha \approx -0.95$, which is sufficiently close to the LFL value of $-1$. 
\begin{figure*}
\begin{center}
\includegraphics[scale=0.8]{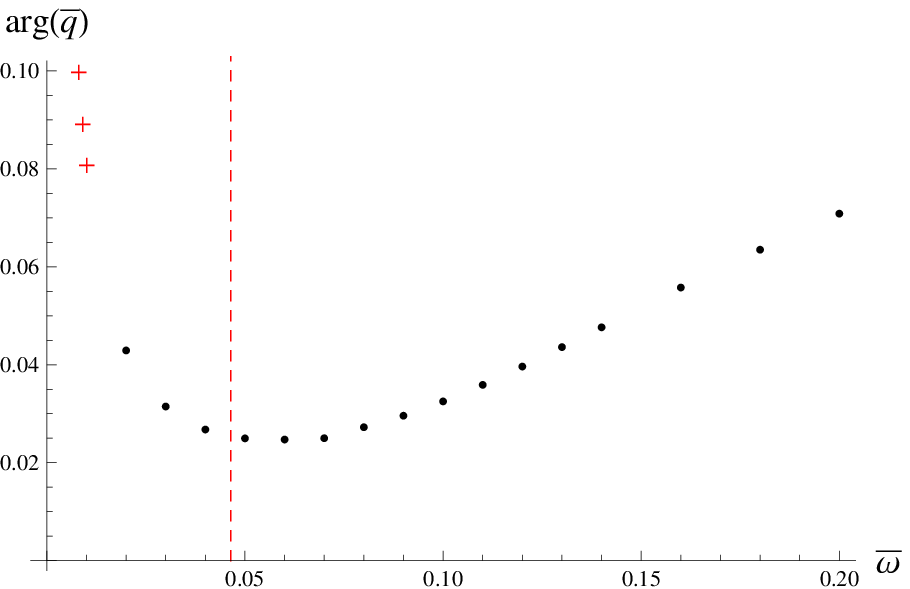}
\includegraphics[scale=0.8]{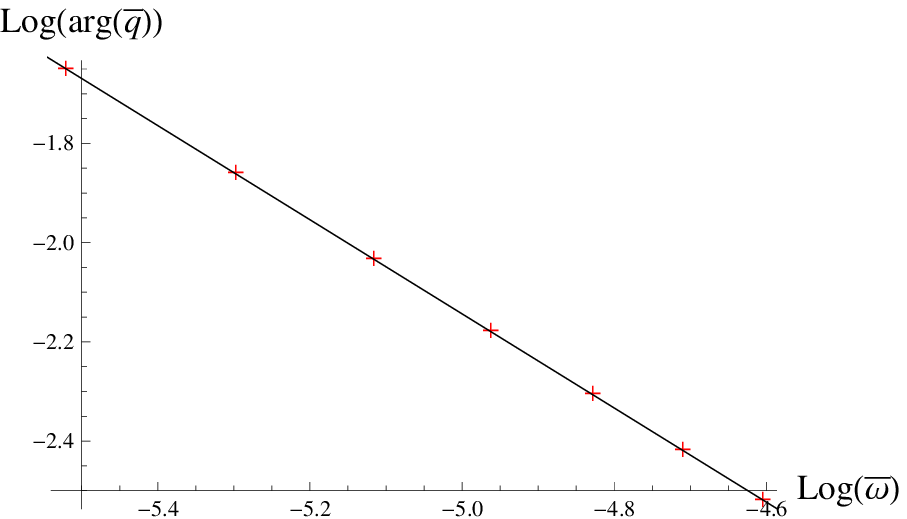}
\caption{The frequency dependence of the D3/D7 acoustic mode when $\mbar=0$, $\dtilde=10^4$ in the collisionless regimes. 
The dots and crosses are our numerical results, the dashed line is 
$\wbar=\pi T/d^{1/3} \sim T/\mu$, 
and the solid line shows the best-fit straight line with gradient $\alpha \approx -0.95$. The points on the left of the left hand plot correspond to the rightmost points on the right hand plot.}
\label{fig:momentumdependence}
\end{center}
\end{figure*} 
Fig.~\ref{fig:momentumdependence} can be compared to regions II and III in Fig.~\ref{argqplot}. As we approach 
the hydrodynamic regime, the power law dependence is lost as expected. We are not able to explore the acoustic
mode in region I due to limitations of the probe brane approximation, as explained in the introduction. The best-fit gradients are the same for non-zero masses $\mbar = 0.76, 1.68$ and so they are not shown here for brevity.

\subsection{The collisionless-hydrodynamic crossover}
\label{sec:chc}

As the temperature is increased further, the zero sound mode becomes less stable, and the system enters 
the hydrodynamic regime characterized by  
\begin{equation}
\label{interval-hr}
 0 \leq \wbar, \qbar \ll \dtilde^{-2/3}\,.
\end{equation}
The zero sound peak in the spectral function broadens and moves to the origin 
(see Fig.~\ref{fig:masslesshighTspectralfunc}).
\begin{figure*}
\begin{center}
\includegraphics[scale=0.7]{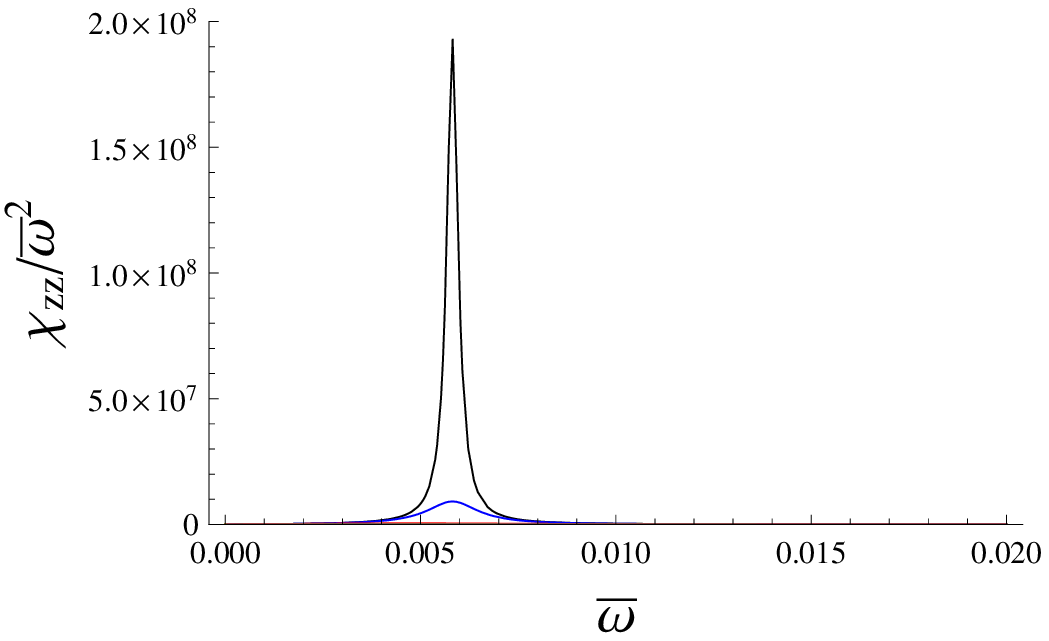}
\includegraphics[scale=0.7]{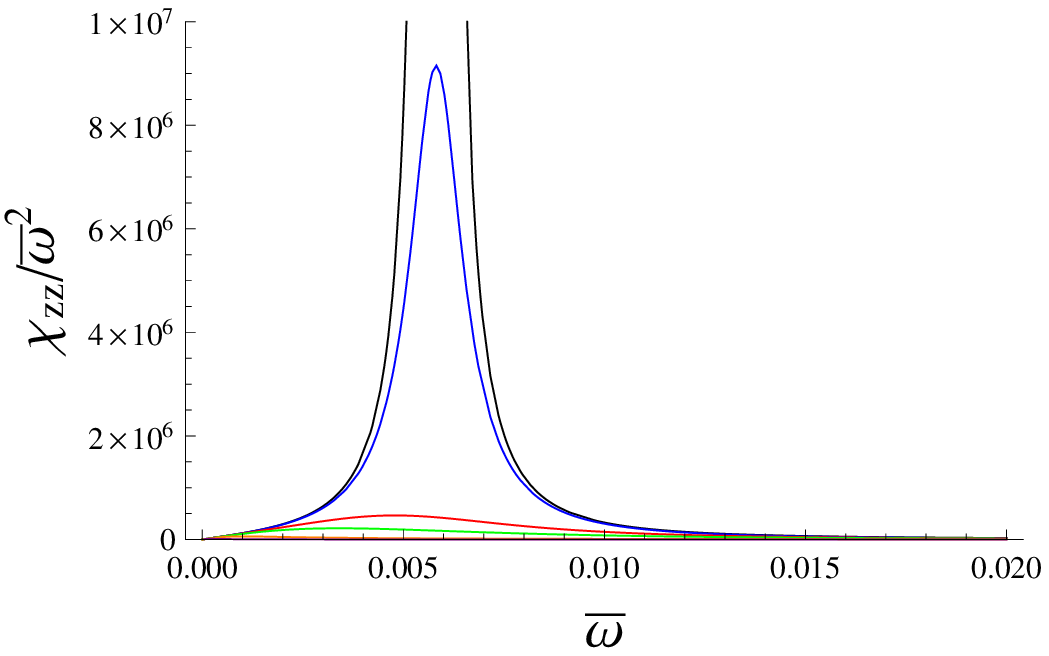}
\includegraphics[scale=0.7]{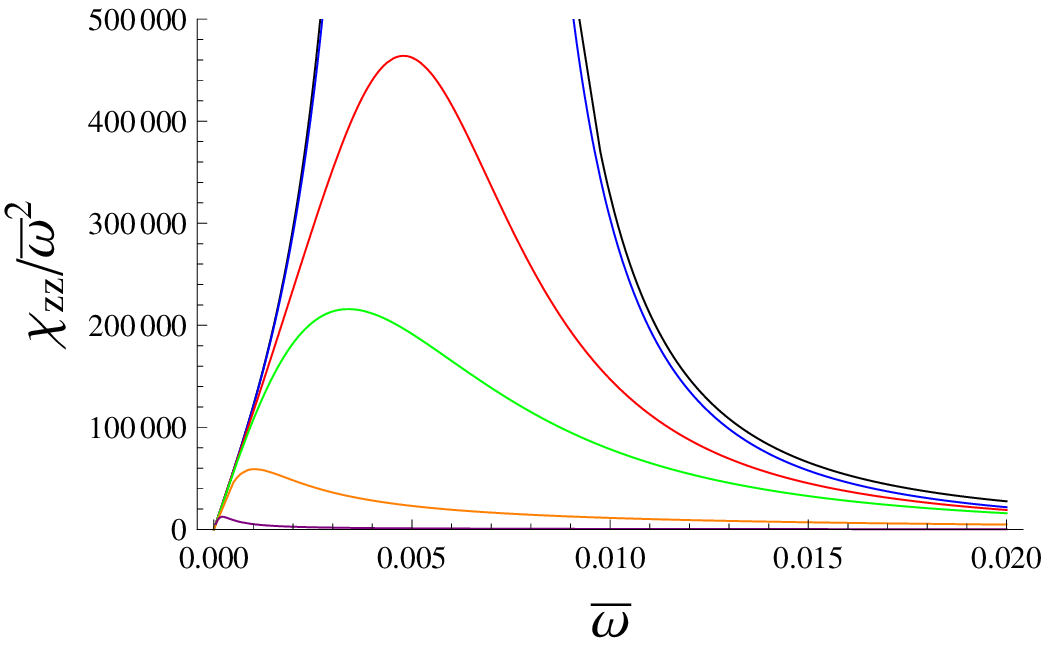}
\includegraphics[scale=0.7]{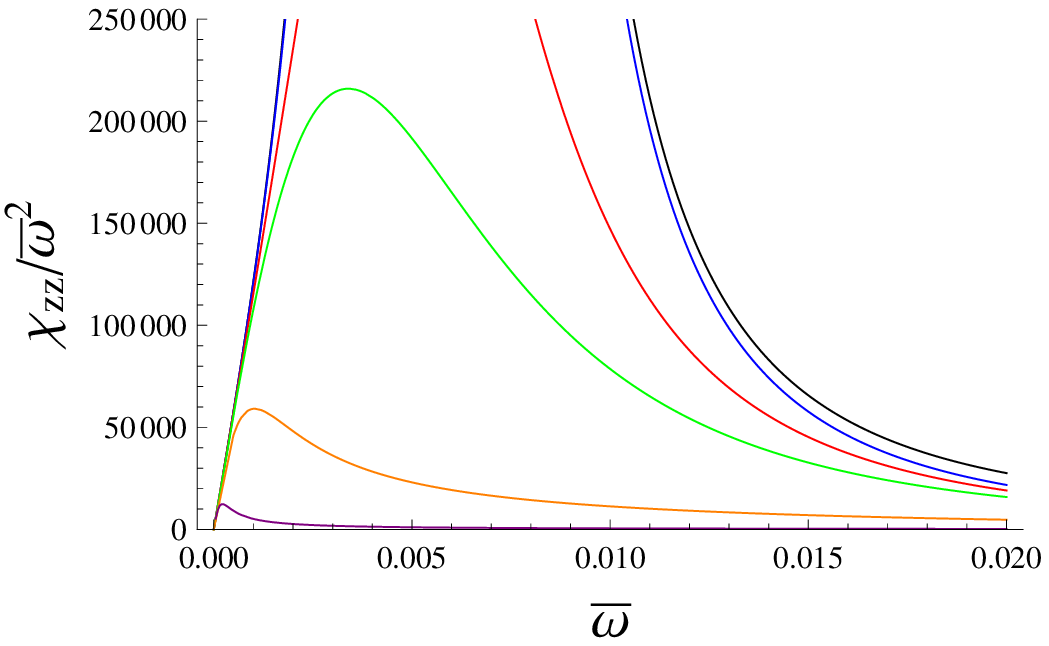}
\includegraphics[scale=0.7]{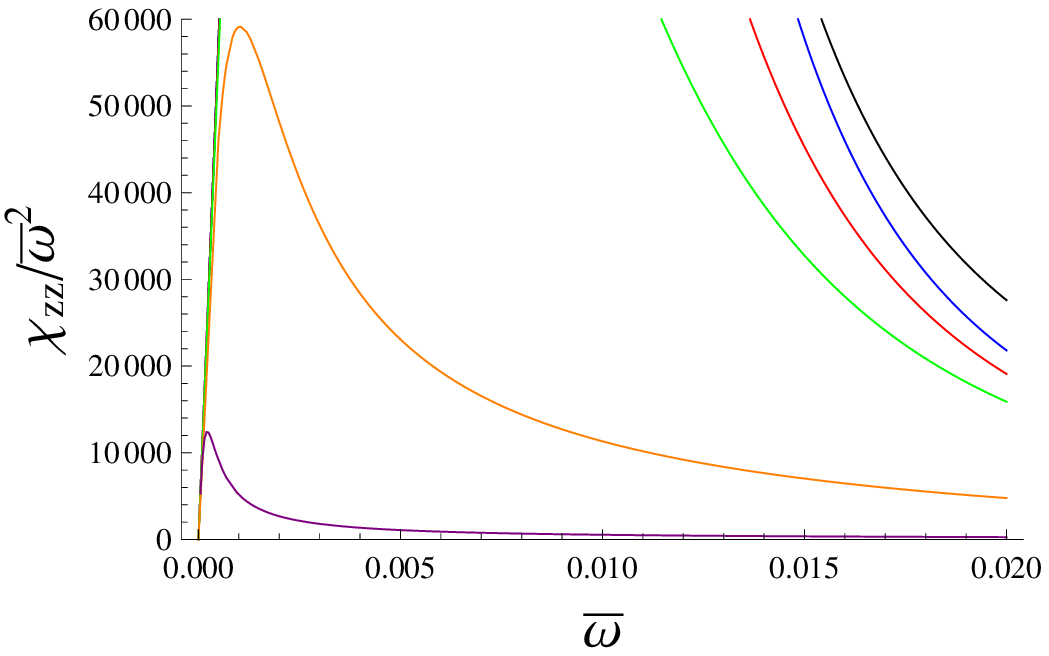}
\includegraphics[scale=0.7]{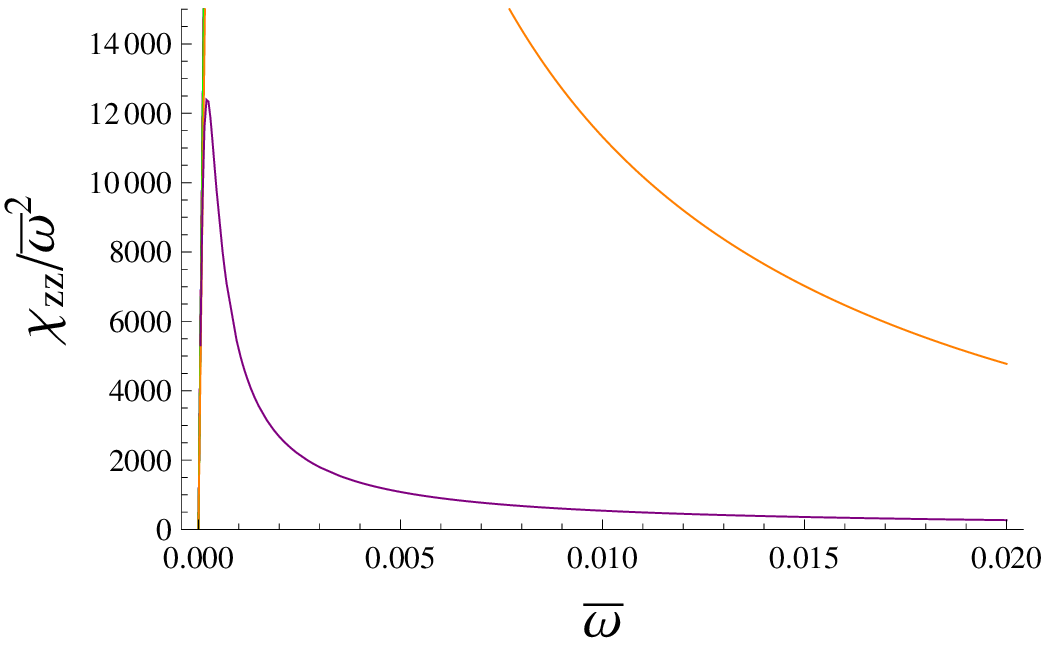}
\caption{The longitudinal spectral functions for $\mbar=0$,
$\qbar=0.01$. Moving from the top left corner to the bottom right corner: each subsequent figure zooms into the 
area of detail in the previous one (note the different scales on the vertical axes). In each figure, 
moving from the tallest peak to the smallest peak corresponds
to raising the temperature: $\dtilde=10^5$ (black), $10^4$ (blue),
$10^3$ (red), $500$ (green), $100$ (orange), $10$ (purple).}
\label{fig:masslesshighTspectralfunc}
\end{center}
\end{figure*} 
The real and imaginary parts of the collective mode's dispersion relation as functions of temperature are shown 
in Fig.~\ref{fig:highTdispersion1} for massless and massive hypermultiplets. The real part decreases with 
increasing temperature until it becomes exactly zero at  $T=T_{cross}$, 
and the mode ceases to propagate. The magnitude of the 
imaginary part increases until $T=T_{cross}$, then decreases again. For $T>T_{cross}$, the mode is purely diffusive, 
approaching at high temperatures the known analytic result (\ref{eq:diffusion}) for the hydrodynamic charge density diffusion mode. The transition occurs at $\qbar \sim \dtilde^{-2/3}$, as we will shortly show.
\begin{figure*}
\begin{center}
\includegraphics[scale=0.82]{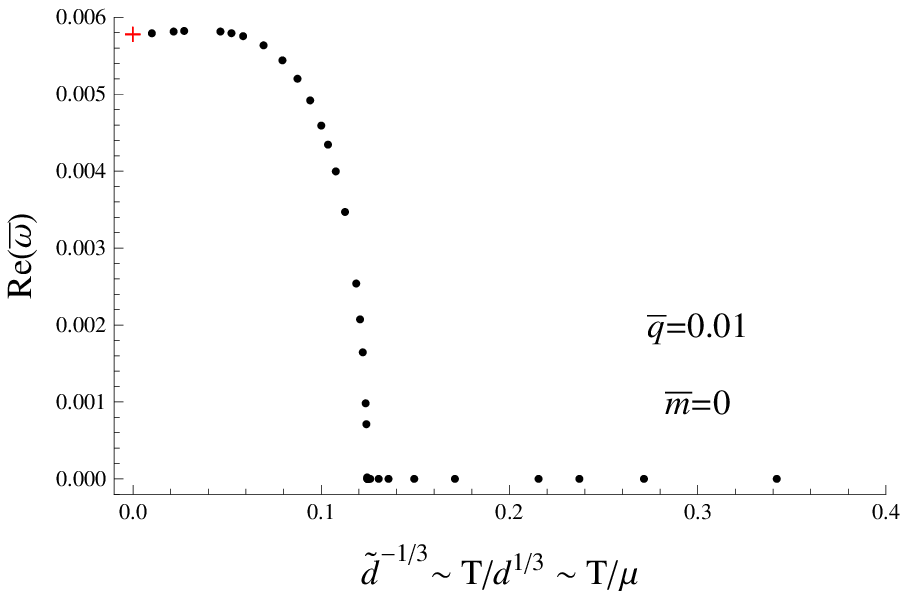}
\includegraphics[scale=0.82]{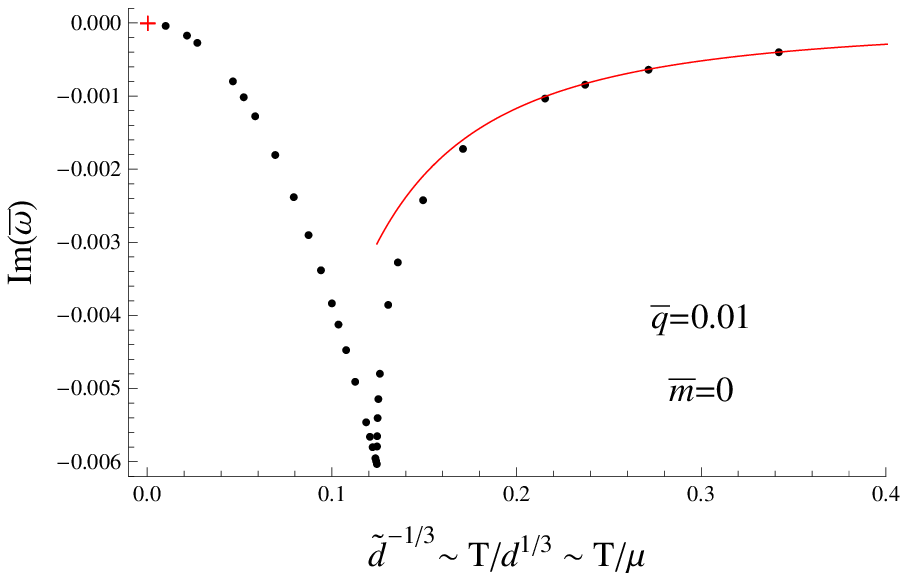}
\includegraphics[scale=0.82]{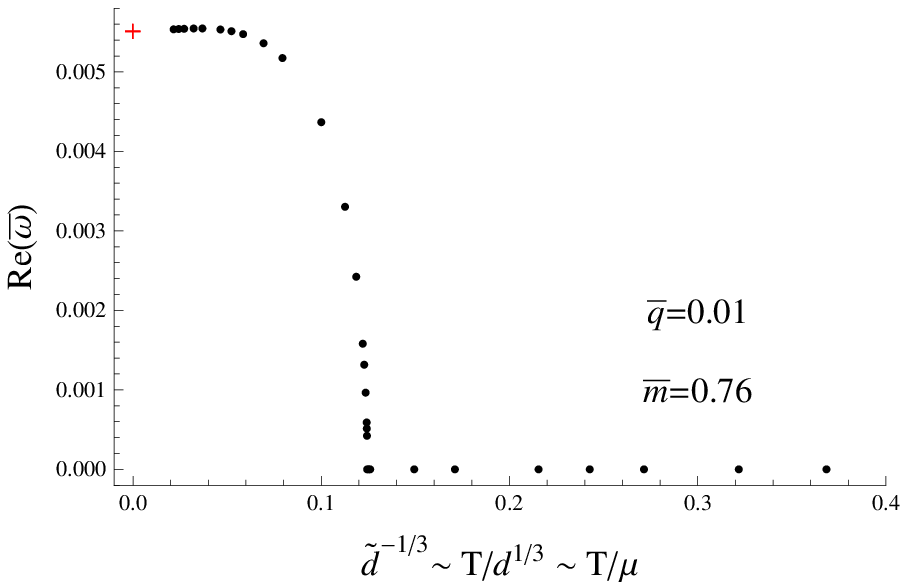}
\includegraphics[scale=0.82]{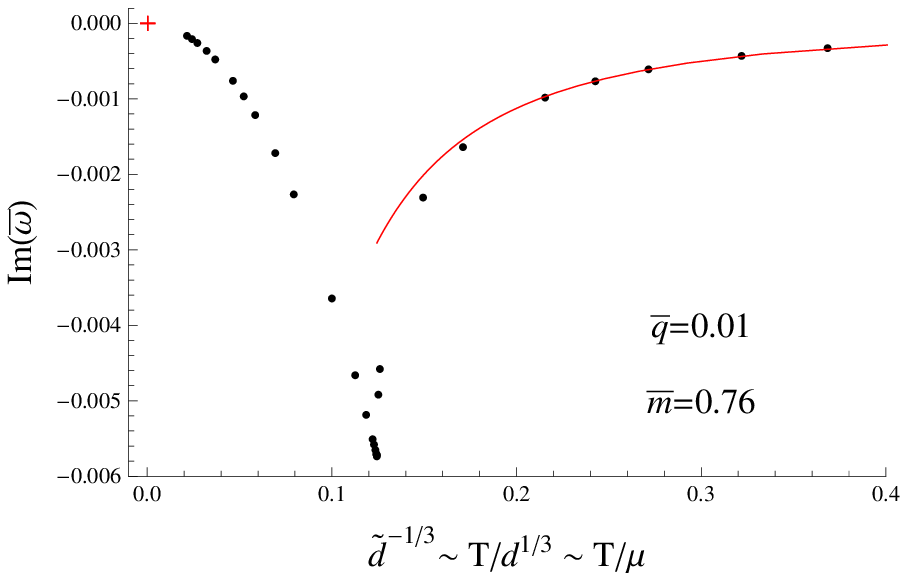}
\includegraphics[scale=0.82]{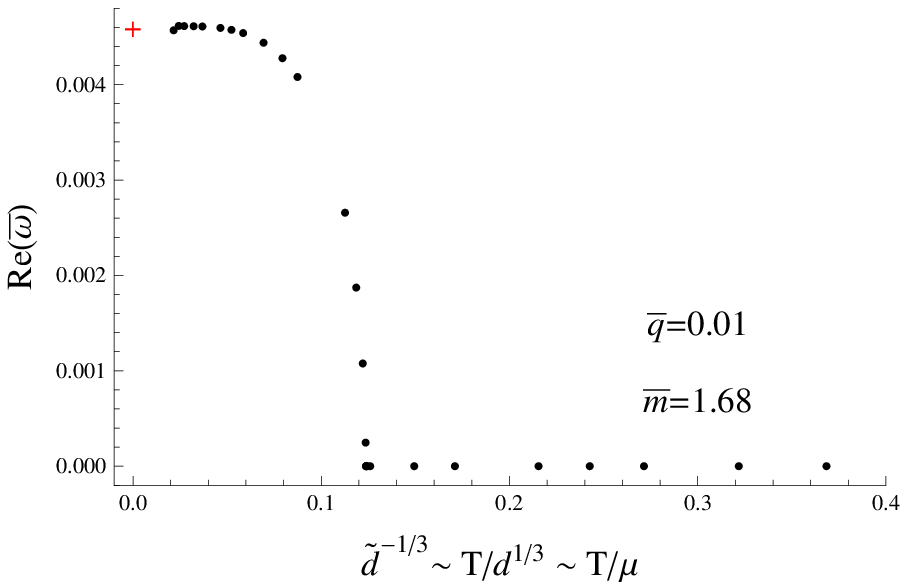}
\includegraphics[scale=0.82]{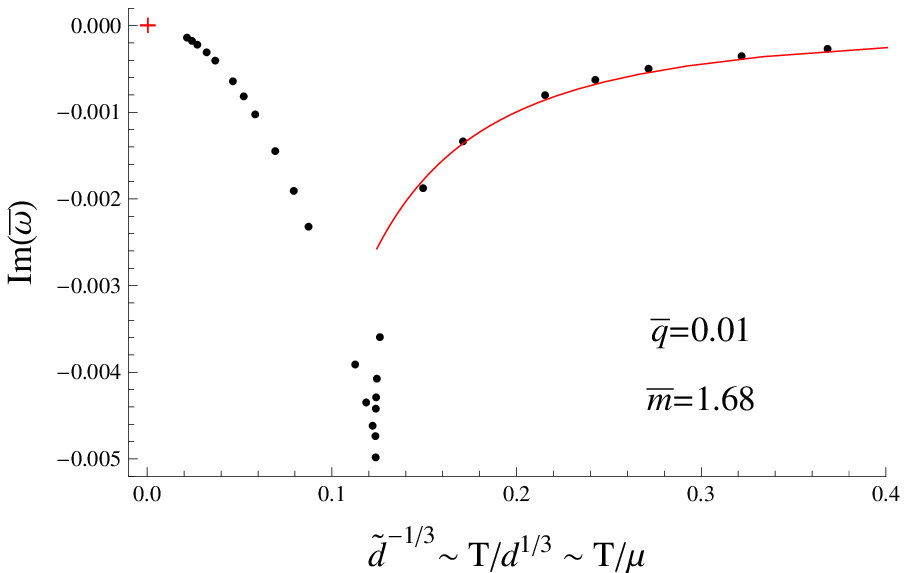}
\caption{The temperature dependence of the real and imaginary parts of the dominant collective mode's 
dispersion relation at $\qbar=0.01$ and $\mbar = 0; 0.76; 1.68$. Dots show the
numerical results at low $T$, the crosses show the $T=0$ zero sound results 
(\ref{eq:zerosound}) and (\ref{eq:massivezerosound}), and the solid lines show the analytic
diffusion result (\ref{eq:diffusion}).}
\label{fig:highTdispersion1}
\end{center}
\end{figure*}

The dynamics of this collisionless-hydrodynamic transition are best exhibited by the motion of the relevant 
poles in the complex frequency plane (see Fig.~\ref{fig:complexpoles1}). 
As the temperature is raised, the two zero sound poles (corresponding to 
$\mbox{Re}\,  \wbar = \pm v_s \qbar$) move deeper into the complex plane, approximately along the circle 
$|\wbar| = \qbar/\sqrt{3}$, until they collide on the imaginary axis and form two purely imaginary poles.
(For the parameters used in Fig.~\ref{fig:complexpoles1}, this happens at $\dtilde = \dtilde_{cross}
 \approx 519$, i.e. in the 
region $\qbar \sim \dtilde^{-2/3}$.) One of these new poles recedes quickly into the complex plane as the 
temperature is raised further, while the other approaches the real axis and becomes the 
hydrodynamic diffusion mode (\ref{eq:diffusion}) at high temperatures. 
This explains the temperature dependence of the collective mode dispersion relation in Fig.~\ref{fig:highTdispersion1} 
and the behavior of the density-density spectral function in Fig.~\ref{fig:masslesshighTspectralfunc}.
At non-zero hypermultiplet mass, the transition is qualitatively similar, although the transition temperature 
decreases slightly with increasing mass ($\dtilde_{cross} \approx 520, 530$ for $\mbar = 0.76, 1.68$, respectively).

The phenomenon in which two poles on the imaginary axis collide and generate two propagating 
modes persists at lower densities as well (in fact, it was first observed in Fig.~3(c) of \cite{Kaminski:2009dh} 
for $\dtilde =2$ which is the highest value of $\dtilde$ studied in that paper), although in the low 
density regime the pole collision occurs deep in the complex plane and the propagating mode is short-lived. 
The pole merger has also been recently observed and 
correctly identified as the hydrodynamic - collisionless transition involving 
the zero sound in a 2+1-dimensional theory in \cite{Bergman:2011rf}. 
\begin{figure}
\begin{center}
\includegraphics[scale=0.75]{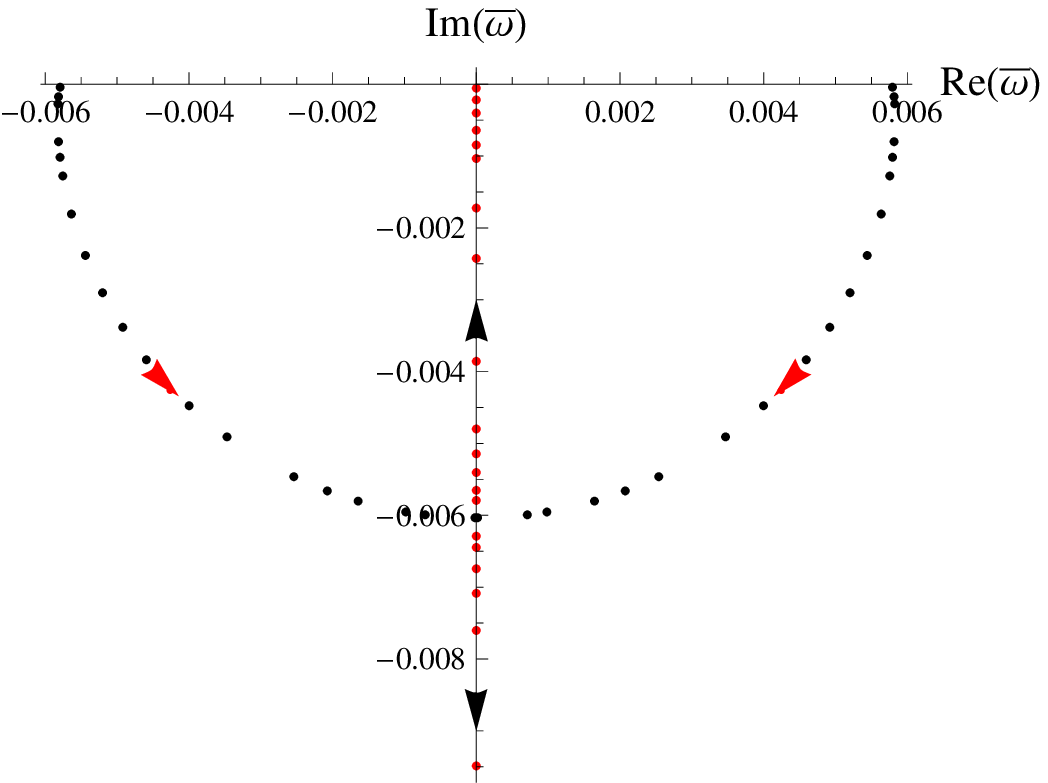}
\caption{The positions of the dominant poles of the density-density correlator 
in the complex frequency plane
at $\mbar=0$, $\qbar=0.01$ as the temperature is
changed. The low temperature limit corresponds to the points furthest from the imaginary axis. 
As the temperature is increased, the points move inwards towards the imaginary axis where 
they collide to form two poles that move up and down the imaginary axis respectively. An animated version of this figure, and the corresponding animations for $\mbar=0.76,1.68$, are available at \url{http://www.physics.ox.ac.uk/users/Davison/D3D7animations.html}.}
\label{fig:complexpoles1}
\end{center}
\end{figure}

In Landau Fermi-liquid theory, the collisionless-hydrodynamic transition occurs when
$\omega\, \tau \sim 1$ and $q\, l_{\text{mfp}} \sim 1$, where $\tau$ and $l_{\text{mfp}}$ are 
the mean free time and mean free path, respectively, of the quasiparticles 
in the vicinity of the Fermi surface supporting the collective mode. 
Defining the collisionless-hydrodynamic transition in the D3/D7 holographic model as 
the event in which the two poles in Fig.~\ref{fig:complexpoles1} merge, we can cast 
the corresponding parameters in the familiar language of the Landau theory by 
introducing an effective $\tau = 1/|\omega_\text{cross}|$ and $l_{\text{mfp}} = 1/q_\text{cross}$,
and computing their temperature dependence. Assuming a simple power-dependence of the form
\begin{equation}
l_{\text{mfp}}\sim\tau\propto d^{-1/3}\left(\frac{T}{d^{1/3}}\right)^{\alpha},
\end{equation}
we expect a plot of $\log(\bar{q}_{cross})$ or $\log(|\bar{\omega}_{cross}|)$ versus
 $\log(\tilde{d}^{1/3}_{cross})$ to yield a straight line of gradient $\alpha$. 
These plots are shown in Fig.~\ref{fig:collisionfrequency} for the massless case.
\begin{figure*}
\begin{center}
\includegraphics[scale=0.64]{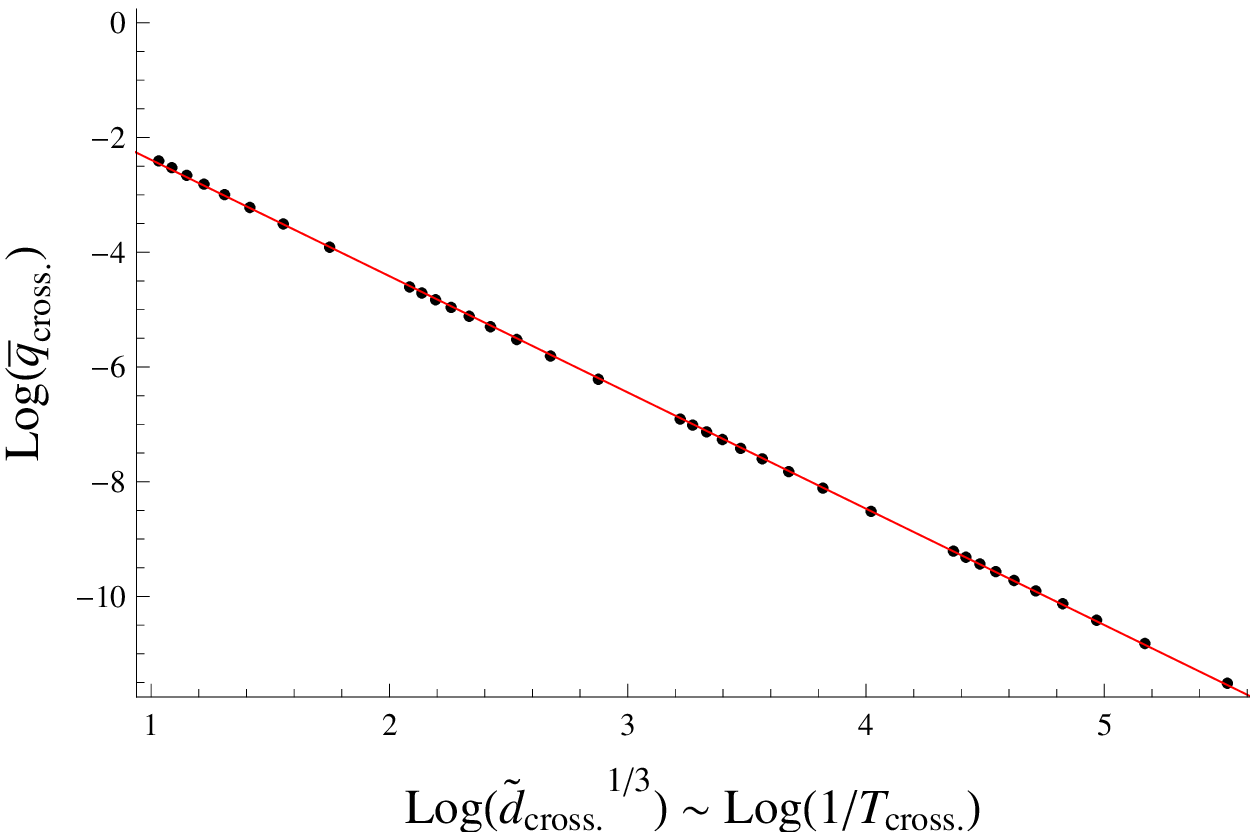}
\includegraphics[scale=0.64]{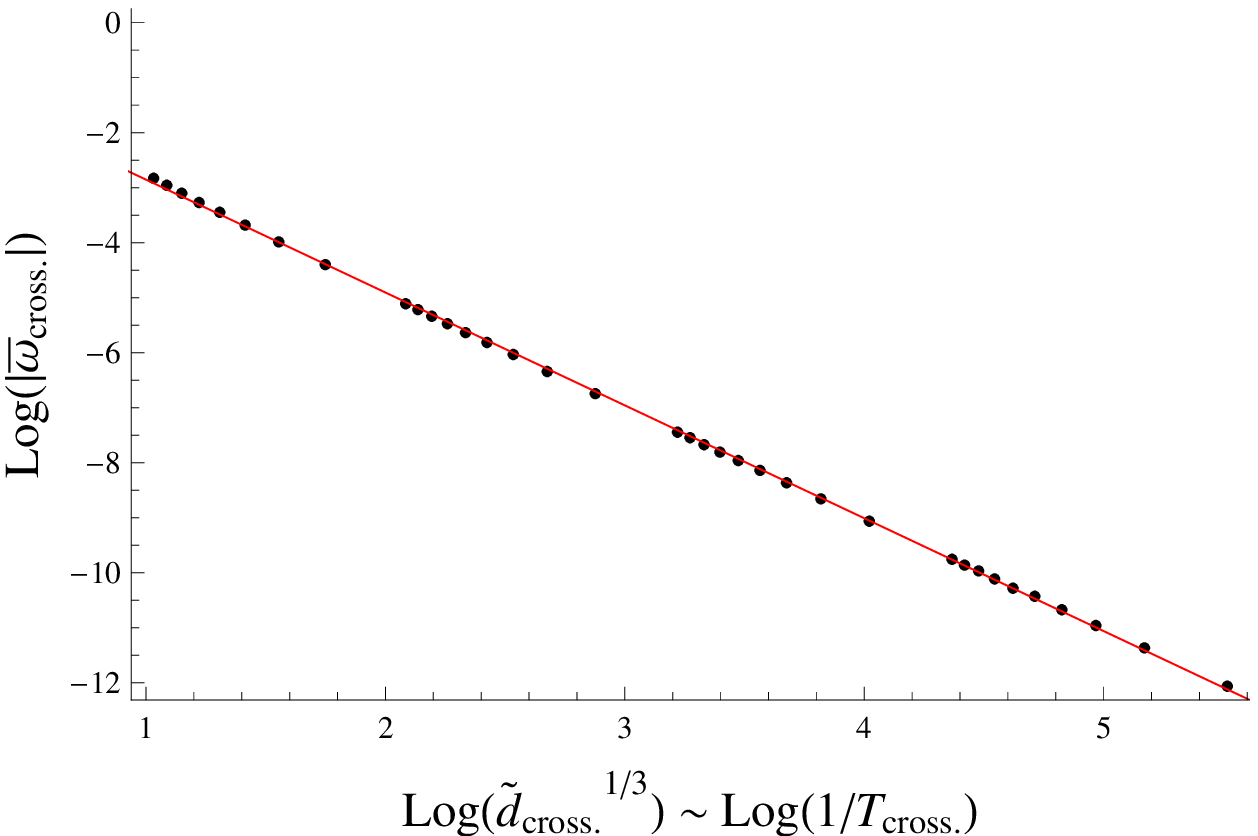}
\caption{The temperature dependence of the collisionless-hydrodynamic crossover value of frequency and 
momentum for $\mbar=0$. The points are our numerical results and the solid lines are the best-fit straight lines which 
both have gradient -2.0.}
\label{fig:collisionfrequency}
\end{center}
\end{figure*} 
The numerical results are clearly consistent with a simple power law dependence. The slope
 of the best-fit straight line is -2.0 in each case, which leads to the result
\begin{equation}
l_\text{mfp} \sim \tau \sim d^{1/3}\, T^{-2}\sim\mu\, T^{-2},
\end{equation}
as in a Landau Fermi-liquid \cite{pinesnozieres} - recall table \ref{tab0}. For non-zero masses $\mbar = 0.76, 1.68$, the best-fit slopes are unchanged and so we do not show the results for brevity.

\section{Summary and discussion}
\label{sec:discussion}

In this paper, we have investigated the properties of the holographic zero sound mode and the fundamental matter diffusion mode in the D3/D7 system at finite temperature and high density. Similarly to the case of an ordinary Landau Fermi-liquid, three regimes 
corresponding to different powers of the parameter $T/\mu \ll 1$ can be identified, 
with the modes having a distinctively different temperature and momentum 
dependence in each regime. In the collisionless quantum regime, the zero sound attenuation is 
essentially temperature-independent and proportional to the square of the momentum whereas 
in the collisionless thermal regime, the attenuation is momentum-independent and scales as the square of the temperature 
(see Table \ref{tab1}, Figs.~\ref{argqplot}, \ref{fig:log-gamma-plots-vertical}
and the corresponding D3/D7 results in 
Figs.~\ref{fig:lowTcorrections}, \ref{fig:momentumdependence}). The crossover transition
between the two regimes is clearly visible in Figs.~\ref{fig:lowTcorrections}, \ref{fig:momentumdependence}. In the hydrodynamic regime of a Landau Fermi-liquid, the acoustic attenuation is proportional to
the square of the momentum and scales as $\sim 1/T^2$. Restricted by the probe brane 
approximation, we are unable to study the acoustic collisionless-hydrodynamic crossover transition in the D3/D7 theory directly.\footnote{See \cite{Bigazzi:2011it} for some recent work on going beyond the probe approximation.} However, we do observe this transition in the density-density correlator as the motion of
the lowest-lying poles in the complex frequency plane leads from the zero 
sound-dominated regime to the diffusion-dominated one as the parameter $T/\mu$ is increased (Fig.~\ref{fig:complexpoles1}). The transition occurs at $\omega, q \sim T^2/\mu$, as if it were supported by Landau Fermi-liquid quasiparticles with a lifetime $\sim 1/T^2$. 
These results do not qualitatively change as the mass of the fundamental matter in the field theory is
varied, at least in the range  $0\leq\mbar\leq1.68$ which we have studied numerically.

The main conclusion of our study is that the holographic zero sound of the D3/D7 system at finite temperature behaves 
exactly as the usual Landau zero sound. This is rather difficult to reconcile, at least within the standard
Landau Fermi-liquid paradigm, with two other properties of the system: the atypical temperature dependence of the specific heat ($c_V \propto T^6/d$ instead of $c_V \propto T d^{2/3}$) and the apparent absence of a singularity when $|\vec{q}|=2q_F$ at zero frequency in the density-density correlator at zero temperature. One may add to these apparent discrepancies a finite zero-temperature limit of the 
entropy density \cite{Karch:2008fa}. Resolution of the above mentioned puzzles may involve
modifications of the gravitational background considered, possibly in the spirit 
of \cite{Hartnoll:2009ns}. On the other hand, the absence of the singularity in the zero frequency density-density correlator at $2k_F$ may have a simpler explanation\footnote{We are indebted to Andrei Parnachev and Pavel Kovtun for valuable discussions on this issue.}: assuming the validity of the Luttinger theorem, 
$n_q \sim k_F^3$, and therefore $q \sim 2 k_F \sim n_q^{1/3} \sim d^{1/3} \lambda^{1/6}$, since $d\sim n_q/\sqrt{\lambda}$. This means that the corresponding $\qbar$ is large, $\qbar \sim \lambda^{1/6}$, 
and is not visible in the correlators in the probe brane limit. It is also possible that the $\sim T^6$
behavior of the specific heat is a genuine property of the considered microscopic model in the approximation described by the probe brane limit, there is no 
Fermi surface, and we are dealing with a new type of quantum liquid. This point of view, initially advocated in  \cite{Karch:2008fa}, is not in contradiction with the findings presented in our paper.

There are other examples of theories with a dual holographic description which support a sound 
mode at $T=0$. It would be interesting to determine whether the results presented here hold (qualitatively) in those cases also. The first class of such theories are probe brane theories \cite{Kulaxizi:2008jx,Hung:2009qk,HoyosBadajoz:2010kd,Lee:2010ez,Bergman:2011rf}. Of particular interest are the D4/D8/$\overline{\text{D8}}$ theory, whose sound mode has a different dispersion relation from Landau Fermi-liquid theory at $T=0$, and the D3/D3 theory which is (1+1)-dimensional and hence to which Landau Fermi-liquid theory does not apply. A qualitatively different holographic theory is the $AdS_4$ Einstein-Maxwell theory at finite density \cite{Liu:2009dm,Faulkner:2009wj,Edalati:2010pn,Belliard:2011qq,Gauntlett:2011mf,Gauntlett:2011wm}. In this theory, non-trivial density-dependent physics is possible via the strong coupling of the charge density to the energy density of the field theory. This is in contrast to probe theories, where the DBI coupling between the charge density gives rise to interesting density-dependent physics, despite the fact that the coupling of this sector to the overall energy density is suppressed via the probe limit. These are very different mechanisms and hence it would be very interesting to see if the low temperature 
sound mode in the Einstein-Maxwell theory behaves similarly to the corresponding sound 
mode in the D3/D7 theory.

Finally, it would be very interesting to generalize the recent approach to zero sound 
proposed in \cite{Nickel:2010pr} to non-zero temperatures.

\acknowledgements
We would like to thank Andy O'Bannon for useful discussions and for bringing 
Fig. 3 of Ref.~\cite{Kaminski:2009dh} to our attention. A.O.S. thanks Johanna Erdmenger, 
Kristan Jensen, Chris Herzog, Pavel Kovtun, Manuela Kulaxizi, 
Andrei Parnachev, Giuseppe Policastro, Mukund Rangamani, Eva Silverstein, Dam Thanh Son and Jan 
Zaanen for helpful discussions and comments on the manuscript, 
and KITP for hospitality during the very last phase of the project. The work of R.A.D. was 
supported by a UK Science and Technology Facilities Council (STFC) studentship. 
The work of A.O.S. was supported, in part, by an STFC Advanced Fellowship.

\appendix
\section{Fluctuations at non-zero hypermultiplet mass}
\label{appendixA}
When the hypermultiplet has a non-zero mass, the bulk
fluctuations of the embedding scalar and the longitudinal gauge field
components are coupled and their solutions are no longer
independent. In the field theory, this corresponds to mixing of the dual
operators. An appropriate  systematic formalism which we follow was introduced in
\cite{Kaminski:2009dh}.
It will be convenient to work with the
gauge-invariant variables
\begin{equation} \varphi\left(u,\wbar\right) =
r_H\tilde{\phi}\left(u,\wbar\right)\,, \qquad 
\Ebar\left(u,\wbar\right) = -i\left[\wbar\tilde{a}_z(u,\wbar)+\qbar\tilde{a}_t(u,\wbar)\right].
\end{equation} 
Note that the definition of $\Ebar$ differs from the massless case
(\ref{eq:gicombo}) by a factor of $-i$. 
The coupled system of equations of motion for the variables $\Ebar$ and $\varphi$ is
\begin{eqnarray}
\label{eq:EOM1}
&\,& \frac{d}{du}\Biggl( 
\frac{f(u)\cos^3\theta}{\sqrt{G(u)}D(u,\wbar)} \Biggl[H(u)\Ebar'(u,\wbar) + 
4i\qbar f(u)u^2\theta'(u)A(u)\varphi'(u,\wbar) \\ \nonumber
&+& 3i\qbar\tan\theta
A(u)G(u)\varphi(u,\wbar)\Biggr]\Biggr)
+\frac{\dtilde^{\frac{2}{3}}\cos^3\theta
H(u)}{4uf(u)\sqrt{G(u)}}\Ebar(u,\wbar) \\ \nonumber 
&+& \frac{\dtilde^{\frac{2}{3}}\cos^3 \theta
uA(u)\theta'(u)}{\sqrt{G(u)}} i \qbar \varphi(u,\wbar) = 0,
\end{eqnarray}
and
\begin{eqnarray}
\label{eq:EOM2}
&\,& 
\frac{d}{du}\Biggl( \frac{f(u)\cos^3\theta}{u\sqrt{G(u)}D(u,\wbar)}\Biggl[ 
\left(\wbar^2-\qbar^2 f(u) B(u) \right)\varphi'(u,\wbar)  \\ \nonumber 
&-& 
4 i \qbar f(u)u^3\theta'(u) A(u) 
\Ebar'(u,\wbar) 
-
3\tan\theta \theta'(u)G(u)\left(\wbar^2-\qbar^2f(u)\right)\varphi(u,\wbar)\Bigr]\Biggr)
\\ \nonumber
&+& \frac{3\cos^2\theta\sin\theta f(u)A(u)\sqrt{G(u)}}{D(u,\wbar)}i\qbar\Ebar'(u,\wbar) \\ \nonumber 
&-& 
\frac{\cos^3\theta
uA(u)\theta'(u)\dtilde^{\frac{2}{3}}}{\sqrt{G(u)}}i\qbar\Ebar(u,\wbar)
+ \frac{3\cos^2\theta\sin\theta
f(u)\theta'(u)\sqrt{G(u)}\left(\wbar^2-\qbar^2f(u)\right)}{uD(u,\wbar)}\varphi'(u,\wbar) \\ \nonumber
&+&
\frac{\cos^3\theta\dtilde^{\frac{2}{3}}\left(\wbar^2-\qbar^2f(u)B(u)\right)}{4u^2f(u)\sqrt{G(u)}}\varphi(u,\wbar)
- \frac{9\cos\theta\sin^2\theta A(u)^2\sqrt{G(u)}\wbar^2}{D(u,\wbar)}\varphi(u,\wbar) \\ \nonumber 
&+& 
\frac{3\left(3\cos^3\theta-2\cos\theta\right)\sqrt{G(u)}}{4u^3}\varphi(u,\wbar)  =0,
\end{eqnarray}
where the coefficients are given by
\begin{eqnarray}
\nonumber
&\,& A(u) = \frac{A_t'(u)}{r_H}\,, \qquad B(u) = 1-4u^3A(u)^2\,, \qquad
 G(u) = 1+4u^2f(u)\theta'(u)^2-4u^3A(u)^2, \\ \nonumber
&\,&  H(u) = 1+4f(u)u^2\theta'(u)^2\,, \qquad D(u,\wbar) = \wbar^2H(u)-\qbar^2f(u)G(u), \nonumber 
\end{eqnarray}
and primes denote derivatives with respect to $u$.

The relevant part of the off-shell action quadratic in longitudinal
fluctuations is
\begin{eqnarray}
\label{eq:actionmassive} 
 S^{(2)}_{\text{long.}} &=& -Nr_H^2\int_0^1 du
\frac{d\omega dq}{\left(2\pi\right)^2}\Biggl\{-\frac{\cos^3\theta f(u)H(u)}{\sqrt{G(u)} D(u,\wbar)}
\, \Ebar'(u,-\wbar) \, \Ebar'(u,\wbar) \\ \nonumber 
&-& 
\frac{8\cos^3\theta f(u)^2u^2\theta'(u)A(u)}{\sqrt{G(u)}D(u,\wbar)} i \qbar \, \Ebar'(u,-\wbar)\, \varphi'(u,\wbar) 
\\ \nonumber 
&-&\frac{\cos^3\theta f(u)\left(\wbar^2-\qbar^2f(u)B(u)\right)}{u\sqrt{G(u)}D(u,\wbar)} \, 
\varphi'(u,-\wbar)\, \varphi'(u,\wbar) 
\\ \nonumber 
&+&\frac{6\cos^2\theta\sin\theta f(u)A(u)\sqrt{G(u)}}{D(u,\wbar)} i \qbar \, \varphi(u,-\wbar) \, 
\Ebar'(u,\wbar) \\ \nonumber
&+&\frac{6\cos^2\theta\sin\theta f(u)\theta'(u)\sqrt{G(u)}\left(\wbar^2-\qbar^2f(u)\right)}{uD(u,\wbar)}
\, \varphi(u,-\wbar)\, \varphi'(u,\wbar) \\ \nonumber
&+& \text{non-derivative terms}\Biggr\},
\end{eqnarray}
where $N=N_fT_{D7}V_{S^3}$. The equations of motion and the action are written in the form that allow one to apply the recipes of
\cite{Kaminski:2009dh} directly. To obtain the retarded Green's functions,
we solve the coupled system of equations (\ref{eq:EOM1}) - (\ref{eq:EOM2})  
with incoming wave boundary conditions at the horizon and combine with the appropriate
 factors from the action (\ref{eq:actionmassive}) as
described in \cite{Kaminski:2009dh}.

\section{Holographic zero sound attenuation at finite hypermultiplet mass and zero temperature}
\label{appendixB}

In this appendix we derive the formula for the zero sound attenuation at finite hypermultiplet mass following the 
approach of Ref.~\cite{Kulaxizi:2008kv}.  Note that our notation is different from that used there. At zero temperature and non-zero mass, it will be convenient to use the coordinate system 
in which the background metric takes the form
\begin{equation}
\nonumber 
ds_{10}^2=\frac{r^2}{R^2}\left(-dt^2+d\vec{x}^2\right)+\frac{R^2}{r^2}\left(d\rho^2+\rho^2ds_{S^3}^2+d\mathcal{R}^2+\mathcal{R}^2d\phi^2\right),
\end{equation}
where $r^2=\rho^2+\mathcal{R}^2$. The background DBI action is given by
\begin{equation}
\nonumber
S_{\text{fund.}}=-N\int_0^\infty d\rho\,  d^4x \, 
\rho^3\sqrt{1+\mathcal{R}'(\rho)^2-A_t'(\rho)^2},
\end{equation}
where $\mathcal{R}$ is the embedding coordinate of the D7-branes. 
The background gauge field and embedding coordinate are determined by two conserved charges $c$ and $d$ via
\begin{equation}
\nonumber
\begin{aligned}
A_t(\rho)&=\frac{R^2}{6}d\left(d^2-c^2\right)^{-\frac{1}{3}}\mathcal{B}\left[\frac{\rho^6}{\rho^6+d^2R^{12}-c^2R^{12}};\frac{1}{6},\frac{1}{3}\right],\\
\mathcal{R}(\rho)&=\frac{R^2}{6}c\left(d^2-c^2\right)^{-\frac{1}{3}}\mathcal{B}\left[\frac{\rho^6}{\rho^6+d^2R^{12}-c^2R^{12}};\frac{1}{6},\frac{1}{3}\right]\,,
\end{aligned}
\end{equation}
where $d$ is the density of fundamental matter and $\mathcal{B}$ is the incomplete beta function.
The mass and chemical potential are given by the asymptotic values
\begin{equation}
\nonumber
A_t\left(\rho\rightarrow\infty\right) = r_H\, \tilde{\mu},\qquad
\mathcal{R}\left(\rho\rightarrow\infty\right) = \frac{r_H}{\sqrt{2}}\, \tilde{m},
\end{equation}
and are related to the constants $d$ and $c$ via
\begin{equation}
\nonumber
c=r_H^3\gamma\frac{\tilde{m}}{\sqrt{2}}\left(\tilde{\mu}^2-\tilde{m}^2/2\right),\qquad
d=r_H^3\gamma\tilde{\mu}\left(\tilde{\mu}^2-\tilde{m}^2/2\right),
\end{equation}
where
\begin{equation}
\nonumber
\gamma=\left(\frac{R^2}{6}\mathcal{B}\left[\frac{1}{6},\frac{1}{3}\right]\right)^{-3}.
\end{equation}
The equations of motion for the fluctuations of the embedding scalar $\delta\mathcal{R}$ and the fluctuations of the gauge-invariant combination $\delta\mathcal{A}\equiv R^2\omega a_z+R^2qa_t$ are 
coupled\footnote{They are given by equations (A.11) of \cite{Kulaxizi:2008kv}, 
with the replacements $E\rightarrow\delta\mathcal{A}$, $\tilde{\xi}\rightarrow\delta\mathcal{R}$, 
$r\rightarrow\rho$, $d\rightarrow dR^6$, $c\rightarrow cR^6$, $L\rightarrow R$, 
$R\rightarrow\mathcal{R}$.}. In the near-horizon limit $\rho\rightarrow0$, the solutions take the form
\begin{equation}
\label{nh-sol}
\delta\mathcal{A} = A\, \rho\, e^{\pm i\Omega/\rho}, \qquad
\delta\mathcal{R} = B\, \rho\,  e^{\pm i\Omega/\rho},
\end{equation}
where $\Omega=R^2\omega\sqrt{1-c^2/d^2}$ and the upper sign in the exponent corresponds to the ingoing boundary condition. In the small frequency limit ($\Omega/\rho\ll1$), the solutions (\ref{nh-sol}) can be expanded as
\begin{equation}
\label{eq:expansion1}
\delta\mathcal{A} = \pm i \Omega  A + A  \rho +\cdots\,, \qquad
\delta\mathcal{R} = \pm i\Omega B + B \rho + \cdots.
\end{equation}

Alternatively, taking first the small frequency limit of the equations of motion, we obtain the 
following solutions near the boundary
\begin{equation}
\nonumber
\delta\mathcal{A} = C_0 + O\left(1/\rho^2\right),\qquad
\delta\mathcal{R} = \tilde{C}_0 + O\left(1/\rho^2\right),
\end{equation}
whereas near the horizon the corresponding solutions are
\begin{equation}
\begin{aligned}
\label{eq:expansion2}
\delta\mathcal{A}&=C_0+b_1C_1+b_2C_2+a_1C_1\rho+a_2C_2\rho,\\
\delta\mathcal{R}&=\tilde{C}_0+\tilde{b}_1C_1+\tilde{b}_2C_2+\tilde{a}_1C_1\rho+\tilde{a}_2C_2\rho,
\end{aligned}
\end{equation}
where the coefficients are given by
\begin{equation}
\nonumber
\begin{aligned}
a_1&= -\frac{c^2q^2+(d^2-c^2)\omega^2}{R^2\left(d^2-c^2\right)^\frac{3}{2}}, \hspace{48.5mm} a_2=\frac{cdq^2}{R^2\left(d^2-c^2\right)^\frac{3}{2}},\\
b_1&=\frac{\Gamma\left(\frac{7}{6}\right)\Gamma\left(\frac{4}{3}\right)\left[\left(3c^2-d^2\right)q^2+3\left(d^2-c^2\right)\omega^2\right]}{\Gamma\left(\frac{1}{2}\right)\left(d^2-c^2\right)^\frac{4}{3}}, \hspace{11mm} b_2=-\frac{2cdq^2\Gamma\left(\frac{7}{6}\right)\Gamma\left(\frac{4}{3}\right)}{\Gamma\left(\frac{1}{2}\right)\left(d^2-c^2\right)^\frac{4}{3}},\\
\tilde{a}_1&=-\frac{cd}{R^6\left(d^2-c^2\right)^\frac{3}{2}}, \hspace{57mm} \tilde{a}_2=\frac{d^2}{R^6\left(d^2-c^2\right)^\frac{3}{2}},\\
\tilde{b}_1&=\frac{2cd\Gamma\left(\frac{7}{6}\right)\Gamma\left(\frac{4}{3}\right)}{R^4\Gamma\left(\frac{1}{2}\right)\left(d^2-c^2\right)^\frac{4}{3}}, \hspace{51mm} \tilde{b}_2=-\frac{\Gamma\left(\frac{7}{6}\right)\Gamma\left(\frac{4}{3}\right)\left(3d^2-c^2\right)}{R^4\Gamma\left(\frac{1}{2}\right)\left(d^2-c^2\right)^\frac{4}{3}}.
\end{aligned}
\end{equation}
The poles of the retarded Green's function correspond to the quasinormal modes of the 
system - these are the solutions which obey ingoing boundary conditions near the horizon 
and vanish near the boundary. Such solutions exist when it is possible to match the expansion (\ref{eq:expansion1}) (with the upper sign) onto the expansion (\ref{eq:expansion2}) with $C_0=\tilde{C}_0=0$. Performing this matching order-by-order in $\rho$, we find that the coefficients $C_1$ and $C_2$ obey the equation
\begin{equation}
\begin{pmatrix}
i\Omega a_1-b_1&i\Omega a_2-b_2\\
i\Omega\tilde{a}_1-\tilde{b}_1&i\Omega\tilde{a}_2-\tilde{b}_2
\end{pmatrix}
\begin{pmatrix}
C_1\\
C_2
\end{pmatrix}
=
\begin{pmatrix}
0\\
0
\end{pmatrix}\,.
\end{equation}
A non-trivial solution exists provided that 
\begin{equation}
\begin{vmatrix}
i\Omega a_1-b_1&i\Omega a_2-b_2\\
i\Omega\tilde{a}_1-\tilde{b}_1&i\Omega\tilde{a}_2-\tilde{b}_2
\end{vmatrix}
=0.
\end{equation}
By substituting an expansion of the form $\omega=c_0+c_1q+c_2q^2+\ldots$, we find that the above determinant vanishes when $\omega$ is given by the dispersion relation 
(\ref{eq:massivezerosound}). Note that this method only finds the lowest energy quasinormal modes, as it involves taking the small $\omega$ limit.

\bibliography{zerosoundv1.bib}

\end{document}